\documentclass[10pt]{article}
\usepackage{bbm}
\usepackage[final]{graphics}
\usepackage{amsmath}
\usepackage{amsfonts,amsbsy}
\usepackage{amssymb}

\def\empile#1\over#2{\mathrel{\mathop{\kern 0pt#1}\limits_{#2}}}
\def\bs{\boldsymbol}
\def\wt#1{\widetilde{#1}}

\newcommand{\slv}{\raise.15ex\hbox{$/$}\kern-.53em\hbox{$v$}}
\newcommand{\slF}{\raise.15ex\hbox{$/$}\kern-.53em\hbox{$F$}}
\newcommand{\slL}{\raise.15ex\hbox{$/$}\kern-.53em\hbox{$L$}}
\newcommand{\slP}{\raise.15ex\hbox{$/$}\kern-.53em\hbox{$P$}}
\newcommand{\slp}{\raise.15ex\hbox{$/$}\kern-.53em\hbox{$p$}}
\newcommand{\slq}{\raise.15ex\hbox{$/$}\kern-.53em\hbox{$q$}}
\newcommand{\slR}{\raise.15ex\hbox{$/$}\kern-.53em\hbox{$R$}}
\newcommand{\slQ}{\raise.15ex\hbox{$/$}\kern-.53em\hbox{$Q$}}
\newcommand{\slK}{\raise.15ex\hbox{$/$}\kern-.53em\hbox{$K$}}
\newcommand{\slk}{\raise.15ex\hbox{$/$}\kern-.53em\hbox{$k$}}
\newcommand{\slD}{\raise.15ex\hbox{$/$}\kern-.73em\hbox{$D$}}
\newcommand{\slC}{\raise.15ex\hbox{$/$}\kern-.53em\hbox{$C$}}
\newcommand{\slA}{\raise.15ex\hbox{$/$}\kern-.53em\hbox{$A$}}
\newcommand{\slSigma}{\raise.15ex\hbox{$/$}\kern-.53em\hbox{$\Sigma$}}
\newcommand{\slpartial}{\raise.15ex\hbox{$/$}\kern-.53em\hbox{$\partial$}}
\newcommand{\slcalP}{\raise.15ex\hbox{$/$}\kern-.63em\hbox{$\cal P$}}

\def\p{{\boldsymbol p}}
\def\q{{\boldsymbol q}}

\def\k{{\boldsymbol k}}

\def\x{{\boldsymbol x}}

\def\v{{\boldsymbol v}}

\def\u{{\boldsymbol u}}

\newcommand{\opt}{\mathbbm{T}}

\date{}

\catcode`\@=11

\newcount\@tempcntc
\def\@citex[#1]#2{\if@filesw\immediate\write\@auxout{\string\citation{#2}}\fi
  \@tempcnta\z@\@tempcntb\m@ne\def\@citea{}\@cite{%
        \@for\@citeb:=#2\do%
    {\@ifundefined{b@\@citeb}%
        {\@citeo\@tempcntb\m@ne\@citea%
                \def\@citea{,\penalty\@m\ }{\bf ?}\@warning%
                {Citation `\@citeb' on page \thepage \space undefined}}%
        {\setbox\z@\hbox{\global\@tempcntc0\csname b@\@citeb\endcsname\relax}%%
     \ifnum\@tempcntc=\z@ \@citeo\@tempcntb\m@ne%
       \@citea\def\@citea{,\penalty\@m}%
       \hbox{\csname b@\@citeb\endcsname}%
     \else%
      \advance\@tempcntb\@ne%
      \ifnum\@tempcntb=\@tempcntc%
      \else\advance\@tempcntb\m@ne\@citeo%
      \@tempcnta\@tempcntc\@tempcntb\@tempcntc\fi\fi}}\@citeo}{#1}}%

\def\@citeo{\ifnum\@tempcnta>\@tempcntb\else\@citea
  \def\@citea{,\penalty\@m}%
  \ifnum\@tempcnta=\@tempcntb\the\@tempcnta\else
   {\advance\@tempcnta\@ne\ifnum\@tempcnta=\@tempcntb \else
\def\@citea{--}\fi
    \advance\@tempcnta\m@ne\the\@tempcnta\@citea\the\@tempcntb}\fi\fi}

\catcode`\@=12

\begin{document}

\title{\bf High energy factorization\\ in nucleus-nucleus collisions\\
II - Multigluon correlations}
\author{Fran\c cois Gelis$^{(1)}$, Tuomas Lappi$^{(2)}$, Raju Venugopalan$^{(3)}$}
\maketitle
\begin{center}
\begin{enumerate}
\item Theory Division, PH-TH, Case C01600, CERN,\\
 CH-1211, Geneva 23, Switzerland
\item Institut de Physique Th\'eorique (URA 2306 du CNRS)\\
  CEA/DSM/Saclay, B\^at. 774\\
  91191, Gif-sur-Yvette Cedex, France
\item  Physics Department, Brookhaven National Laboratory\\
  Upton, NY-11973, USA
\end{enumerate}
\end{center}

\maketitle

\begin{abstract}
  We extend previous results (arXiv: 0804.2630 [hep-ph]) on
  factorization in high-energy nucleus-nucleus collisions by computing
  the inclusive multigluon spectrum to next-to-leading order.  The
  factorization formula is strictly valid for multigluon emission in a
  slice of rapidity of width $\Delta Y\leq \alpha_s^{-1}$. Our results
  shows that often neglected disconnected graphs dominate the
  inclusive multigluon spectrum, and are crucial in order to achieve
  factorization for this quantity. These results provide a dynamical
  framework for the Glasma flux tube picture of the striking ``ridge''-like
  correlation seen in heavy ion collisions.
\end{abstract}

\section{Introduction}

In a recent work, henceforth referred to as Paper I~\cite{GelisLV3},
we investigated the formal basis for the application of the Color
Glass Condensate (CGC)
framework~\cite{McLerV1,McLerV2,McLerV3,McLer1,IancuLM3,IancuV1,GelisLV2} to the collision of
two high energy nuclei.  In Paper I, we focused on the formalism to
compute the single gluon inclusive spectrum in the Leading Log $x$
approximation. The main result of Paper I is a proof that terms
containing leading logarithms of $1/x_{1,2}$ that arise in all order
loop corrections to this spectrum can be factorized in the
distributions of color sources $W[\rho_{1,2}]$ in each of the two 
nuclei,
evolved with the JIMWLK
equation~\cite{JalilKMW1,JalilKLW1,JalilKLW2,JalilKLW3,JalilKLW4,IancuLM1,IancuLM2,FerreILM1}
from the beam rapidity to the rapidity of the measured gluon. One
obtains for the single inclusive gluon distribution the result
\begin{equation}
\left<
\frac{dN}{d^3\p}\right>_{_{\rm LLog}}
= \int [D{{\rho}}_1] [D{{\rho}}_2]\;
W_{Y_{\rm beam}-Y}[{\rho}_1]\, W_{Y_{\rm beam}+Y}[{\rho}_2]\;
\left.\frac{dN}{d^3\p}\right|_{_{\rm LO}} \; .
\label{eq:single-gluon}
\end{equation}
The $W$ functionals are universal properties of the nuclear
wavefunctions at high energies and (in analogy to the parton
distribution functions of collinear factorization) can be extracted
from deep inelastic scattering or proton-nucleus scattering
experiments off nuclei.  The inclusive single gluon spectrum
${(dN/d^3\p)}_{_{\rm LO}}$ that appears under the integral in the
right hand side is the Leading Order spectrum corresponding to one
configuration of the sources $\rho_{1,2}$ -- it is obtained by solving
the classical Yang-Mills equations for this fixed distribution of
sources. This factorization theorem allows for considerable predictive
power by relating measurements in a variety of scattering processes.
It should be particularly useful at the LHC, where the rapidity reach
in proton-nucleus and nucleus-nucleus collisions will be considerable
and the effects of energy evolution of the distribution of color sources
clearly visible.

The derivation of the factorized expression in
eq.~(\ref{eq:single-gluon}) relied on two essential steps:
\begin{enumerate}
\item The 1-loop corrections to the gluon spectrum can, in the leading
logarithm approximation,  be expressed as the action of a certain linear 
operator  on the leading order spectrum\footnote{
See the  discussion after eqs.~(40-41) and at the end of section 3.5 in 
\cite{GelisLV3}.},
\item This operator acting on the initial color fields on the
  light-cone is, again in the leading  log approximation,
 the JIMWLK Hamiltonian.
\end{enumerate}
In the present paper, we will show that a straightforward
generalization of the first of these two steps is sufficient to extend
our factorization result to inclusive {\it multigluon spectra} when
all the measured gluons are located in a rapidity region of maximal
width $\Delta Y\lesssim \alpha_s^{-1}$.

The paper is organized as follows. In section \ref{sec:genfunc}, we
define a generating functional for multiparticle production in
nucleus-nucleus collisions. This extends to the QCD case our previous
results~\cite{GelisV2,GelisV3} for a similar object introduced for a
$\phi^3$ theory.  We discuss key features of this generating
functional and develop a diagrammatic interpretation of this object.
We show how (at leading order) its first derivative can be expressed
in terms of classical solutions of the Yang-Mills equations that obey
both advanced and retarded boundary conditions. In section
\ref{sec:2gluon}, we consider in detail the inclusive two-gluon
spectrum. We obtain an expression of this spectrum at next-to-leading
order (NLO) using the previously defined generating functional.  We
end the section by showing that the leading logs of $1/x_{1,2}$ in
this quantity can be factorized in the distributions of incoming color
sources, provided the rapidity separation between the two gluons is
small enough. We show that our formalism gives rise to the Glasma flux
tube picture~\cite{DumitGMV1}, which has been suggested as a mechanism
to describe the ridge-like structure observed in heavy ion collisions
at RHIC~\cite{Adamsa4,Wang1,Adamsa5,Adarea1,Wosie1}.  In section
\ref{sec:multi}, we generalize this factorization result to the case
of the inclusive $n$-gluon spectrum. Knowing all the moments defines
the complete probability distribution. We demonstrate how the leading
logarithmic corrections to the multiplicity distribution can be
factorized into the JIMWLK evolution of the sources. We end with a
brief summary.  The three appendices are devoted to the more technical
aspects of our discussion.

\section{Generating functional}

In Paper I, we developed the tools for studying at LO and NLO the
single inclusive gluon spectrum in AA collisions in the CGC framework.
Our goal is to generalize these techniques to obtain similar results
for the $n$-gluon spectrum. Towards that purpose, we will define in
this section a generating functional for $n$-gluon production, discuss
its properties and develop a diagrammatic interpretation. We then
discuss the LO computation of the first derivative of this object in
terms of solutions of classical Yang-Mills equations with both
retarded and advanced boundary conditions.

\label{sec:genfunc}
\subsection{Definition and properties}
We define the generating functional as 
\begin{equation}
{\cal F}[z(\p)]\equiv
\sum_{n=0}^\infty
\frac{1}{n!}
\int\limits_{\p_1}\cdots\int\limits_{\p_n}
z(\p_1)\cdots z(\p_n)\,\Big|
\big<\p_1\cdots \p_n{}_{\rm out}\big|0{}_{\rm in}\big>
\Big|^2\; ,
\label{eq:F-def}
\end{equation}
where we use the following compact notation for phase-space
integrals\footnote{Whenever the integrand contains $p_0$ in such
  integrals, it should be replaced by the positive on-shell energy
  $p_0=|\p|$.},
\begin{equation}
\int\limits_\p\cdots\equiv\int\frac{d^3\p}{(2\pi)^3 2E_\p}\cdots\; .
\end{equation}
In this definition, $z(\p)$ is an arbitrary function over the 1-gluon
phase-space. The matrix element squared that appears in the right hand
side is implicitly summed over the polarizations and colors of the
produced gluons. Note that in this section, we consider the external
current $J^\mu$ coupled to the gauge field to be fixed. This is the
case in the Color Glass Condensate (CGC) framework~\cite{GelisLV2}
where the fixed sources represent the large $x$ light cone color
charge densities in the nuclear wavefunctions. We will address the
issue of averaging over the external color sources later in this
paper.

The generating functional generalizes the {\sl generating function}
$F(z)$ we introduced in ref.~\cite{GelisV2}. This previously defined
function is simply obtained as
\begin{equation}
{\cal F}[z(\p)\equiv z^*]=F(z^*)\; ,
\end{equation}
when  $z(\p)$  is a constant $z^*$. 
Another obvious property of ${\cal F}[z(\p)]$ is 
\begin{equation}
{\cal F}[z(\p)\equiv 1]=1\; .
\end{equation}
which is a consequence of the fact that the theory is unitary.

The generating functional encapsulates the entire information content
of the nuclear collision within the CGC framework.  Indeed, if ${\cal
  F}[z(\p)]$ were known, one could use it to build an event generator
for the early Glasma~\cite{LappiM1,GelisV3,GelisLV2} stage of
nucleus-nucleus collisions. In particular, one can compute the
inclusive multigluon spectra. For instance, the single
inclusive\footnote{Note that setting $z(\p)$ to zero instead, after
  taking the functional derivative, one obtains the differential
  probability for producing {\sl exactly one} gluon in the collision,
  \begin{equation*}
    \frac{dP_1}{d^3\p}=
    \left.\frac{\delta{\cal F}[z]}{\delta z(\p)}\right|_{z\equiv 0}\; .
  \end{equation*}
  } 
gluon spectrum is obtained as
\begin{equation}
\frac{dN}{d^3\p}
=
\left.\frac{\delta{\cal F}[z]}{\delta z(\p)}\right|_{z\equiv 1}
\; .
\label{eq:1gluon}
\end{equation}
Likewise, the inclusive 2-gluon spectrum is obtained by
differentiating ${\cal F}[z]$ twice,
\begin{equation}
\frac{d N_2}{d^3\p d^3\q}=
\left.\frac{\delta^2{\cal F}[z]}{\delta z(\p)\delta z(\q)}\right|_{z\equiv 1}\; ,
\label{eq:2gluon}
\end{equation}
where the integral over $\p$ and $\q$ on the left hand side of this
expression is the average value of $N(N-1)$.
Physically, this quantity, in an event, corresponds
to a histogram of all pairs of {\sl distinct} gluons with momenta
$(\p,\q)$. We will discuss the average over all such events later.
Eqs.~(\ref{eq:1gluon}) and (\ref{eq:2gluon}) are the two simplest
examples of the use of this generating functional, but in principle
one can derive from it any observable that is related to the
distribution of gluons produced in the collision.
Eq.~(\ref{eq:2gluon}) can be generalized to
\begin{equation}
\frac{d^n N_n}{d^3\p_1\cdots d^3\p_n}=
\left.\frac{\delta^n{\cal F}[z]}{\delta z(\p_1)\cdots\delta z(\p_n)}\right|_{z\equiv 1}\; ,
\label{eq:ngluon}
\end{equation}
for the inclusive $n$-gluon spectrum. Note that the l.h.s, integrated
over the n-particle phase space, is normalized to the average value of
$N (N-1) \cdots (N-n+1)$.

{}From eq.~(\ref{eq:ngluon}), it is possible to represent the generating
functional ${\cal F}[z]$ as 
\begin{eqnarray}
{\cal F}[z(\p)]=\sum_{n=0}^{\infty}\frac{1}{n!}
\int \Big[
\prod_{i=1}^n
d^3\p_i\;
(z(\p_i)-1)\Big]
\;
\frac{d^n N_n}{d^3\p_1\cdots d^3\p_n}\; .
\label{eq:rep1}
\end{eqnarray}
This formula will later be the basis of our strategy to obtain an
expression for ${\cal F}[z]$ at Leading Log. We will first obtain
Leading Log expressions for the $n$-gluon spectra\footnote{With the 
  important limitation that the $n$ gluons all sit in a rapidity
  slice of width $\Delta Y\lesssim \alpha_s^{-1}$.}, and will show
that the infinite sum in eq.~(\ref{eq:rep1}) leads to a very simple
expression.

Once we know ${\cal F}[z]$ (with a given accuracy), one can use the
fact that its Taylor coefficients at $z(\p)=0$ are the differential
probabilities for producing a fixed number of particles\footnote{Note
  that there is no $1/n!$ in this formula. A quick way to convince
  oneself that this is correct is to set $z(\p)=1$; the integrals over
  the momenta $\p_i$ give the total probabilities $P_n$, which add up
  to unity.},
\begin{eqnarray}
{\cal F}[z(\p)]=\sum_{n=0}^{\infty}
\int \Big[
\prod_{i=1}^n
d^3\p_i\;
z(\p_i)\Big]
\;
\frac{d^n P_n}{d^3\p_1\cdots d^3\p_n}\; .
\label{eq:rep2}
\end{eqnarray}
{}From this second representation of ${\cal F}[z]$, one can extract from
${\cal F}[z]$ detailed information about the distribution of produced gluons.

\subsection{Diagrammatic interpretation of ${\cal F}[z]$}
In order to see what are the diagrams that contribute to ${\cal F}[z]$, let
us first define
\begin{equation}
{\cal D}\equiv\int\limits_\p{\cal D}_\p\; ,
\end{equation}
with
\begin{equation}
{\cal D}_\p\equiv
\sum_\lambda
\epsilon^\mu_\lambda(\p)\epsilon^\nu_\lambda(\p)^*
\int d^4x\; d^4y\; e^{ip\cdot(x-y)}\;
\square_x\square_y\;
\frac{\delta}{\delta J^\mu_+(x)}\frac{\delta}{\delta J^\nu_-(y)}\; .
\label{eq:Dp}
\end{equation}
This operator has already been introduced in \cite{GelisV2,GelisV3} to
write $P_n$ in terms of vacuum diagrams.  The only difference here is
that we extend its definition to the case of vector particles and QCD.
The sum over the gluon polarizations $\lambda$ spans the two physical
polarization states. By mimicking the manipulations performed for
scalar fields, one can prove that
\begin{equation}
{\cal F}[z(\p)]=\exp\left[\int\limits_\p z(\p)\;{\cal D}_\p
\right]\,
\left.
e^{iV[J^\mu_+]}\;e^{-iV^*[J^\mu_-]}
\right|_{J^\mu_+=J^\mu_-=J^\mu}\; ,
\end{equation}
where $iV[J^\mu]$ is the sum of the connected vacuum diagrams
evaluated with the external current $J^\mu$. It is easy to check that
all the formulas we previously obtained in \cite{GelisV2,GelisV3}
for $P_n$ or for the generating function $F(z)$ are all particular
cases of this formula. 

{}From the interpretation of the operator ${\cal D}$ as an operator that
makes cuts through vacuum diagrams, we see that the functional ${\cal
  F}[z(\p)]$ is the sum of all the cut vacuum diagrams (connected or
not) in which every cut propagator with momentum $\p$ is weighted by
$z(\p)$. Let us call $i{\cal W}[J^\mu_+,J^\mu_-;z]$ the sum of all
such {\sl connected} diagrams (before the currents $J^\mu_+$ and
$J^\mu_-$ are set equal to the physical value $J^\mu$)~:
\begin{equation}
  e^{i{\cal W}[J^\mu_+,J^\mu_-;z]}
  \equiv
  \exp\left[\int\limits_\p z(\p)\;{\cal D}_\p
  \right]\,
  e^{iV[J^\mu_+]}\;e^{-iV^*[J^\mu_-]}\; .
\end{equation}

It is useful to compute the first derivative of ${\cal F}[z(\p)]$ with
respect to $z(\p)$,
\begin{equation}
\frac{\delta {\cal F}[z]}{\delta z(\p)}
=
\left.
\frac{1}{(2\pi)^3 2E_\p}\;{\cal D}_\p\;e^{i{\cal W}[J^\mu_+,J^\mu_-;z]}\right|_{J^\mu_+=J^\mu_-=J^\mu}\; .
\end{equation}
Performing explicitly the derivatives contained in eq.~(\ref{eq:Dp}),
this can be rewritten as
\begin{eqnarray}
\frac{\delta {\cal F}[z]}{\delta z(\p)}
&=&
\frac{1}{(2\pi)^3 2E_\p}
\sum_\lambda
\epsilon^\mu_\lambda(\p)\epsilon^\nu_\lambda(\p)^*
\int d^4x \; d^4y\; e^{ip\cdot(x-y)}\;
\square_x\square_y\nonumber\\
&&\times
\left[
\frac{\delta i{\cal W}}{\delta J^\mu_+(x)}
\frac{\delta i{\cal W}}{\delta J^\nu_-(y)}
+
\frac{\delta^2 i{\cal W}}{\delta J^\mu_+(x)\delta J^\nu_-(y)}
\right]
\left.
e^{i{\cal W}[J^\mu_+,J^\mu_-;z]}\right|_{J^\mu_+=J^\mu_-=J^\mu}\; .
\end{eqnarray}
The final exponential in this formula is nothing but ${\cal F}[z]$
itself. Therefore, we can write
\begin{eqnarray}
\frac{\delta \ln {\cal F}[z]}{\delta z(\p)}
&=&
\frac{1}{(2\pi)^3 2E_\p}
\sum_\lambda
\epsilon^\mu_\lambda(\p)\epsilon^\nu_\lambda(\p)^*
\int d^4x d^4y\; e^{ip\cdot(x-y)}\;
\square_x\square_y\nonumber\\
&&\quad\times
\left[
\frac{\delta i{\cal W}}{\delta J^\mu_+(x)}
\frac{\delta i{\cal W}}{\delta J^\nu_-(y)}
+
\frac{\delta^2 i{\cal W}}{\delta J^\mu_+(x)\delta J^\nu_-(y)}
\right]_{J^\mu_+=J^\mu_-=J^\mu}
\; .
\label{eq:dF}
\end{eqnarray}
This formula tells us that this quantity is made up of only connected
diagrams since $i{\cal W}$ is a sum of connected diagrams.
We also observe that this formula is very similar to the formula for
the single inclusive particle spectrum with one very important
difference: the function $z(\p)$ is not set to 1 at the end, and
therefore appears as a multiplicative factor attached to each cut
propagator.

\subsection{$\delta\ln{\cal F}[z]/\delta z(\p)$ at leading order}
Let us now show that, in the regime of strong external color sources,
the expression in eq.~(\ref{eq:dF}) can be expressed at leading order (LO) 
in terms of classical solutions of the Yang-Mills equations.

First of all, note that the first derivatives $\delta {\cal W}/\delta
J^\mu_\pm$ are of order\footnote{Because ${\cal W}$ is the sum of
  connected vacuum graphs, in the presence of external sources
  $J^\mu_\pm\sim g^{-1}$, ${\cal W}\sim g^{-2}$.}  $g^{-1}$, while the
second derivative $\delta^2 {\cal W}/\delta J^\mu_+\delta J^\nu_-$ is
order $g^0$. Thus the first term, composed of the product of two first
derivatives, is the leading one. The second term begins to contribute
only at next-to-leading order (NLO).  At LO, we can thus write
\begin{eqnarray}
&&
\left.
\frac{\delta \ln {\cal F}[z]}{\delta z(\p)}
\right|_{_{\rm LO}}
=
\frac{1}{(2\pi)^3 2E_\p}
\sum_\lambda
\epsilon_{\lambda\mu}(\p)\epsilon_{\lambda\nu}(\p)
\int d^4x d^4y\; e^{ip\cdot(x-y)}\;
\nonumber\\
&&\qquad\qquad\qquad\qquad\qquad\qquad\qquad
\times\,
\square_x\square_y
{\cal A}^\mu_+(x){\cal A}^\nu_-(y)
\; ,
\label{eq:dF_LO}
\end{eqnarray}
where we denote\footnote{
${\cal A}_\pm^\mu(x)$ depends on function $z(\p)$ as well but we have omitted it from the notation to 
keep notations compact.}
\begin{equation}
{\cal A}^\mu_\epsilon(x)\empile{\equiv}\over{\mbox{tree}}
\left.\frac{\delta i{\cal W}}{\delta J^\mu_\epsilon(x)}
\right|_{J^\mu_+=J^\mu_-=J^\mu}\; .
\end{equation}
The ``tree''  here means we keep only tree diagrams in the
expansion of $\delta {\cal W}/\delta J^\mu_\pm$ that defines
${\cal A}^\mu_\epsilon$. 

All the arguments developed  to compute the generating function $F(z)$ at
leading order \cite{GelisV3} can be extended trivially to the present
situation, and one obtains the following results~:
\begin{itemize}
\item ${\cal A}^\mu_\epsilon$ is a solution of the classical
  Yang-Mills equations,
\begin{equation}
\big[{\cal D}_\mu,{\cal F}^{\mu\nu}\big]=J^\nu\; ,
\label{eq:EOM}
\end{equation}

\item If one decomposes ${\cal A}^\mu_\epsilon(x)$ in  Fourier modes,
\begin{equation}
{\cal A}^\mu_\epsilon(x)
\equiv
\sum_{\lambda,a}
\int\limits_\p
\left\{
f_\epsilon^{(+)}(x_0;\p\lambda a)\,a^{0\mu}_{-\p\lambda a}(x)
+
f_\epsilon^{(-)}(x_0;\p\lambda a)\,a^{0\mu}_{+\p\lambda a}(x)
\right\}\, ,
\label{eq:fourier}
\end{equation}
with $a^{0\mu}_{\pm\p\lambda a}(x)\equiv \epsilon^\mu_\lambda(\p)T^a
e^{\pm i p\cdot x}$, the boundary conditions o\-be\-yed by the
classical field ${\cal A}^\mu_\epsilon(x)$ can be expressed as simple
constraints on the Fourier coefficients\footnote{The derivation of
  this result is analogous to the scalar case discussed in detail in
  section 4.2 of Ref.~\cite{GelisV2}.},
\begin{eqnarray}
&&
f_+^{(+)}(-\infty;\p\lambda a)=f_-^{(-)}(-\infty;\p\lambda a)=0\; ,
\nonumber\\
&&
f_-^{(+)}(+\infty;\p\lambda a)=z(\p)\; f_+^{(+)}(+\infty;\p\lambda a)\; ,
\nonumber\\
&&
f_+^{(-)}(+\infty;\p\lambda a)=z(\p)\; f_-^{(-)}(+\infty;\p\lambda a)\; .
\label{eq:bc}
\end{eqnarray}
\end{itemize}
We see that the dependence of the classical fields ${\cal A}^\mu_\pm$
on the function $z(\p)$ comes entirely from the boundary
conditions\footnote{\label{foot:retarded}Note that, when $z(\p)\equiv
  1$, the boundary conditions in eqs.~(\ref{eq:bc}) become
\begin{eqnarray*}
  &&
f_+^{(+)}(-\infty;\p\lambda a)=f_-^{(-)}(-\infty;\p\lambda a)=0\; ,
\nonumber\\
&&
f_-^{(+)}(+\infty;\p\lambda a)=f_+^{(+)}(+\infty;\p\lambda a)\; ,\quad
f_+^{(-)}(+\infty;\p\lambda a)=f_-^{(-)}(+\infty;\p\lambda a)\; .
\end{eqnarray*}
The two conditions at $x^0=+\infty$ imply that ${\cal A}_+(x)={\cal
  A}_-(x)$ everywhere. The two conditions at $x^0=-\infty$ then imply
that $\lim_{x^0\to -\infty}{\cal A}_\pm(x)=0$. Therefore, when
$z(\p)\equiv 1$, the two classical fields ${\cal A}^\mu_\pm$ become
identical to the retarded classical field with a vanishing initial
condition in the remote past, and eq.~(\ref{eq:dF_LO}) gives the
single inclusive gluon spectrum as expected.}, since the Yang-Mills
equations themselves do not explicitly contain $z(\p)$.  In terms of
the Fourier coefficients $f_\pm^{(\pm)}$, eq.~(\ref{eq:dF_LO}) reads
\begin{equation}
\left.
\frac{\delta \ln {\cal F}[z]}{\delta z(\p)}
\right|_{_{\rm LO}}
=
\frac{1}{(2\pi)^3 2E_\p}
\sum_{\lambda,a}
f_+^{(+)}(+\infty;\p\lambda a)\,f_-^{(-)}(+\infty;\p\lambda a)\; .
\label{eq:dF-LO1}
\end{equation}
Note that it depends only on the Fourier coefficients of the fields at
$x^0=+\infty$.

Eqs.~(\ref{eq:EOM}), (\ref{eq:bc}) and (\ref{eq:dF-LO1}) do not
provide a practical way to obtain the LO generating functional ${\cal
  F}[z(\p)]$ because the solutions depend on boundary conditions at
both $\pm \infty$. It is not known at present how to solve Yang--Mills
equations with simultaneous advanced and retarded boundary conditions.
Nevertheless, the procedure outlined here provides a powerful
theoretical tool to compute other quantities, that can be obtained as
derivatives of the generating functional. A concrete illustration of
this strategy is revealed in the case of the 2-gluon spectrum in the
following section.

\section{Two-gluon inclusive spectrum}
\label{sec:2gluon}
In this section, we will specialize our discussion of the generating
functional in the previous section to the 2-gluon inclusive spectrum
at LO and NLO. We will demonstrate that, just as in the case of the
single gluon spectrum discussed in Paper I, the leading logarithm
contributions that arise at NLO can be absorbed in the JIMWLK
wave functionals of the two nuclei, provided the rapidity separation
between the two gluons is small enough. As in Paper I, one obtains a
factorized expression for the leading log 2-gluon inclusive spectrum.
In the following section, this result will be extended to multigluon
spectra.

\subsection{Leading Order}
\label{subsec:2gluon-LO}
The inclusive 2-gluon spectrum is obtained by taking the second
derivative of the generating functional ${\cal F}[z]$, and by setting
the functions $z(\p)$ and $z(\q)$ to unity afterwards (see
eq.~(\ref{eq:2gluon})).  Alternately, it is easy to obtain this
derivative from the derivative of $\ln{\cal F}[z]$.  We get
\begin{equation}
\frac{d^2 N_2}{d^3\p\; d^3\q}=\left.
\frac{\delta\ln{\cal F}[z]}{\delta z(\p)}
\frac{\delta\ln{\cal F}[z]}{\delta z(\q)}
+
\frac{\delta^2\ln{\cal F}[z]}{\delta z(\p)\delta z(\q)}
\right|_{z(\p),z(\q)\equiv 1}
\label{eq:2gluon-1}
\; .
\end{equation}
The first term is simply the product of two single gluon spectra (see
eq.~(\ref{eq:1gluon})), and therefore corresponds to the disconnected
(independent) production of a gluon of momentum $\p$ and a gluon of
momentum $\q$.  In contrast, because $\ln{\cal F}[z]$ contains only
connected diagrams, the second term corresponds to the two gluons
being produced in the same graph. Note that these expressions
correspond to the 2-gluon spectrum for a fixed configuration of the
external sources $\rho_{1,2}$. When we average over these sources,
some graphs that were disconnected prior to averaging become
connected. Therefore, even the first term in eq.~(\ref{eq:2gluon-1})
can lead to correlations in the measured 2-gluon spectrum.

The two terms in this expression do not begin at the same order in
$g^2$. In our power counting,
\begin{equation}
\ln{\cal F}[z]=\frac{1}{g^2}\Big[c_0+c_1\,g^2+c_2\,g^4+\cdots\Big]\; .
\end{equation}
This implies that the first term in eq.~(\ref{eq:2gluon-1}) is of
order $g^{-4}$, while the second term is of order $g^{-2}$ only. For
the 2-gluon spectrum, ``leading order'' therefore means $g^{-4}$, and
we simply have\footnote{One should keep in mind therefore that ``LO'' 
corresponds to different powers of $g^2$ for the single and double inclusive
  gluon spectra.}
\begin{equation}
\left.
\frac{d^2 N_2}{d^3\p d^3\q}\right|_{_{\rm LO}}
=
\left.
\frac{d N}{d^3\p}\right|_{_{\rm LO}}
\left.
\frac{d N}{d^3\q}\right|_{_{\rm LO}}\; .
\label{eq:N2-LO}
\end{equation}
No new computations are necessary here because we know how to express
the single gluon spectrum at LO in terms of classical solutions of the
Yang-Mills equations with retarded boundary conditions. Note that at
this order the $-N$ term contributing to $N_2$ is subleading relative
to the $N^2$ contribution because it starts only at the order
$g^{-2}$ therefore does not  appear on the right hand side of
eq.~(\ref{eq:N2-LO}) which is of order $g^{-4}$.

\subsection{Next to Leading Order - I}
We shall now study the inclusive 2-gluon spectrum at NLO--the
contribution at order $g^{-2}$ in our power counting. At this order,
the tree level contribution to second term in eq.~(\ref{eq:2gluon-1})
must be included. We can write therefore
\begin{equation}
\left.
\frac{d^2 N_2}{d^3\p d^3\q}\right|_{_{\rm NLO}}
=
\left.
\frac{d N}{d^3\p}\right|_{_{\rm NLO}}
\left.
\frac{d N}{d^3\q}\right|_{_{\rm LO}}
+
\left.
\frac{d N}{d^3\p}\right|_{_{\rm LO}}
\left.
\frac{d N}{d^3\q}\right|_{_{\rm NLO}}
+
\left.\frac{\delta^2\ln{\cal F}[z]}{\delta z(\p)\delta z(\q)}\right|_{_{\rm LO}}
\; .
\label{eq:N2-NLO}
\end{equation}
The first two terms again do not require a new computation because we
studied in great detail the single gluon spectrum at NLO in Paper
I~\cite{GelisLV3}. In particular, we recall here the previously
derived formula
\begin{eqnarray}
&&
\left.
\frac{d N}{d^3\p}\right|_{_{\rm NLO}}
=
\Bigg[\,
\underbrace{
\int\limits_{\Sigma}
d^3\vec\u
\big[\beta\cdot\opt_\u\big]
}_{{\cal L}_1}
\nonumber\\
&&
\qquad+
\underbrace{
\frac{1}{2}\sum_{\lambda,a}
\int\limits_\k
\int\limits_{\Sigma}
d^3\vec\u\,d^3\vec\v\;
\big[a_{-\k \lambda a}\cdot\opt_\u\big]
\big[a_{+\k \lambda a}\cdot\opt_\v\big]
}_{{\cal L}_2}
\Bigg]
\left.
\frac{d N}{d^3\p}\right|_{_{\rm LO}}
\nonumber\\
&&\qquad+\Delta N_{_{\rm NLO}}(\p)\; .
\label{eq:O-NLO1}
\end{eqnarray}
In this formula,
$a_{\pm\k\lambda a}$ denotes small field fluctuations that propagate
over the classical field ${\cal A}$. The subscripts indicate that
these fluctuations begin in the remote past as plane waves of momentum
$\pm k$, polarization $\lambda$ and color $a$. Similarly, $\beta$ is
also a small field fluctuation propagating on top of ${\cal A}$, but
this fluctuation has a vanishing initial condition in the past and is
driven by a non zero source term.  $\Sigma$ is a surface on which the
initial value of the classical fields are defined, and $d^3\vec\u$ is
the measure on this surface.  The operator ${\mathbbm T}_\u$ is the
generator of translations of the initial field at the point
$\u\in\Sigma$.  $\Delta N_{_{\rm NLO}}(p)$ is a term contributing to
the full expression. It will not be explicited further because it does
not contain a leading logarithmic contribution--see the discussion of
this term in \cite{GelisLV3}. Because we are interested here in these
leading log contributions, this term will be dropped in all further
equations in this paper.

At this point, we can rewrite the first two terms of the r.h.s. of
eq.~(\ref{eq:N2-NLO}) as 
\begin{eqnarray}
&&
\left.
\frac{d N}{d^3\p}\right|_{_{\rm NLO}}
\left.
\frac{d N}{d^3\q}\right|_{_{\rm LO}}
+
\left.
\frac{d N}{d^3\p}\right|_{_{\rm LO}}
\left.
\frac{d N}{d^3\q}\right|_{_{\rm NLO}}
=
\Big[{\cal L}_1+{\cal L}_2
\Big]_{\rm disc}\;
\left.\frac{d N}{d^3\p}\right|_{_{\rm LO}}
\left.\frac{d N}{d^3\q}\right|_{_{\rm LO}}\; ,
\nonumber\\
&&
\label{eq:N2-NLO1}
\end{eqnarray}
where the subscript ``disc'' added to the operator between the square
brackets indicates that when the combination ${\mathbbm T}_\u{\mathbbm
  T}_\v$ in ${\cal L}_2$ acts on the product $(dN/d^3\p)(dN/d^3\q)$,
we keep only the terms where the two ${\mathbbm T}$'s act on the same
factor\footnote{$\big[{\cal L}_2\big]_{\rm disc}\,AB=\big[{\cal L}_2
  A\big]B+A\big[{\cal L}_2 B\big]$.}. The subscript here reminds us
that these terms are disconnected contributions that are the product
of a function of $\p$ and a function of $\q$.

\subsection{Next to Leading Order - II}
The third term of eq.~(\ref{eq:N2-NLO}), involving the second
derivative of the log of the generating functional, is new and will be
computed here. Fortunately, we need this term only at leading
order--i.e. ${\cal O}(g^{-2})$. Therefore, our starting point in
evaluating this term is eq.~(\ref{eq:dF-LO1}).  Differentiating this
equation with respect to $z(\q)$, we obtain
\begin{eqnarray}
&&
\left.\frac{\delta^2\ln{\cal F}[z]}{\delta z(\p)\delta z(\q)}\right|_{_{\rm LO}}
=\frac{1}{(2\pi)^3 2E_\p}
\nonumber\\
&&
\!\!\!\!\!\!
\times
\sum_{\lambda,a}
\Big[
\frac{\delta f_+^{(+)}(+\infty;\p\lambda a)}{\delta z(\q)}\,
f_-^{(-)}(+\infty;\p\lambda a)
+
f_+^{(+)}(+\infty;\p\lambda a)\,
\frac{\delta f_-^{(-)}(+\infty;\p\lambda a)}{\delta z(\q)}
\Big]\; .
\nonumber\\
&&
\label{eq:ddF-LO1}
\end{eqnarray}
Further, differentiating eq.~(\ref{eq:fourier}) with respect to
$z(\q)$, one observes that the quantities ${\delta
  f_\epsilon^{(\pm)}(+\infty;\p\lambda a)}/{\delta z(\q)}$ are the
Fourier coefficients of the field
\begin{equation}
b_{\epsilon,\q}^\mu(x)\equiv 
\frac{\delta{\cal A}_\epsilon^\mu(x)}{\delta z(\q)}\; 
\end{equation}
at $x^0=+\infty$. The equation of motion obeyed by this object can be
obtained by differentiating, with respect to $z(\q)$, the equation of
motion for ${\cal A}^\mu_\epsilon$. In order to do this, it is useful
to start from the Yang-Mills equations written in a form that
separates explicitly the kinetic and interaction terms\footnote{Note
  that the differentiation with respect to $z(\q)$ does not modify the
  gauge fixing condition, provided it is linear. Thus,
  $b_{\epsilon,\q}^\mu$ obeys the same gauge condition as ${\cal
    A}^\mu_\epsilon$.} as
\begin{equation}
\Big[
\square_x g_{\mu}{}^\nu-\partial_{x\mu}\partial_x^\nu
\Big]{\cal A}_\epsilon^\mu(x)
-
\frac{\partial U({\cal A}_\epsilon)}{\partial {\cal A}_{\epsilon,\nu}(x)}
=J^\nu_\epsilon\; ,
\label{eq:YM}
\end{equation}
where $U({\cal A})$ is the Yang-Mills potential in a gauge with a
linear gauge condition. Differentiating this equation with respect to
$z(\q)$, we get
\begin{equation}
\Big[
\square_x g_{\mu}{}^{\nu}-\partial_{x\mu}\partial_x^\nu
-
\frac{\partial U({\cal A}_\epsilon)}{\partial {\cal A}_{\epsilon\nu}(x)\partial {\cal A}_\epsilon^\mu(x)}
\Big]\,b_{\epsilon,\q}^\mu(x)
=0\; .
\label{eq:fluct}
\end{equation}
In other words, $b_{\epsilon,\q}^\mu(x)$ obeys the equation of motion
of small fluctuations propagating on top of the classical field ${\cal
  A}_\epsilon$. The boundary conditions necessary in order to fully
determine $b_{\epsilon,\q}^\mu(x)$ are easily obtained by
differentiating the eqs.~(\ref{eq:bc}) with respect to $z(\q)$:
\begin{eqnarray}
&&
b_{+,\q}^{(+)}(-\infty;\p\lambda a)=b_{-,\q}^{(-)}(-\infty;\p\lambda a)=0\; ,
\nonumber\\
&&
b_{-,\q}^{(+)}(+\infty;\p\lambda a)=z(\p)\; b_{+,\q}^{(+)}(+\infty;\p\lambda a)
+\delta(\p-\q)f_+^{(+)}(+\infty;\p\lambda a)
\; ,
\nonumber\\
&&
b_{+,\q}^{(-)}(+\infty;\p\lambda a)=z(\p)\; b_{-,\q}^{(-)}(+\infty;\p\lambda a)
+\delta(\p-\q)f_-^{(-)}(+\infty;\p\lambda a)
\; ,
\label{eq:bc-a}
\end{eqnarray}
where we have introduced the obvious notation 
\begin{equation}
b_{\epsilon,\q}^{(\eta)}(x^0;\p\lambda a)\equiv 
\frac{\delta f_{\epsilon}^{(\eta)}(x^0;\p\lambda a)}{\delta z(\q)}
\label{eq:Fourier-b}
\end{equation}
for the Fourier
coefficients of $b_{\epsilon,\q}^\mu$. We see that we have {\sl non homogeneous boundary conditions}, which
will lead to a non zero $b_{\epsilon,\q}^\mu$ despite the fact that
this fluctuation obeys an homogeneous equation of motion. Note also
that at this point we can safely set $z(\p)=1$ since we do not need
to differentiate with respect to $z(\p)$ again. This leads to the 
simplification that when $z(\p)=1$, the classical fields ${\cal A}_+^\mu$
and ${\cal A}_-^\mu$ become identical--as can be checked from their
boundary conditions (see footnote \ref{foot:retarded}). In fact, their common
value is nothing but the classical field that vanishes when $x^0\to
-\infty$.  We will simply denote by ${\cal A}^\mu$ the common value of
these two fields and $f^{(\pm)}(x^0;\p\lambda a)$ its Fourier
coefficients.

Obviously, eqs.~(\ref{eq:bc-a}) are not simple retarded boundary
conditions. Our task is now to relate the fluctuations
$b_{\epsilon,\q}^\mu$ and their Fourier coefficients to 
fluctuations that satisfy simple retarded boundary conditions. In
order to achieve this, let us again use the small field fluctuations
$a_{\pm\k\lambda a}^\mu$. They obey the equation of
motion~(\ref{eq:fluct}), and the boundary conditions
\begin{equation}
a_{\pm\k\lambda a}^\mu(x)\empile{=}\over{x^0\to-\infty}\epsilon^\mu_\lambda(k) 
T^a e^{\pm i k\cdot x}\; ,
\end{equation}
Note that the fields $a_{\pm\k\lambda a}^{0\mu}$ introduced earlier
are the analogue of the $a_{\pm\k\lambda a}^{\mu}$ in the absence
of a background field. From this definition, $a_{+\k\lambda a}$ has
only negative energy components at $x^0\to-\infty$, while
$a_{-\k\lambda a}$ has only positive energy components in this limit.
Moreover, the fluctuations $a_{\pm\k\lambda a}^\mu$ provide a complete
basis for the small field fluctuations that obey eq.~(\ref{eq:fluct}).
{}From the boundary conditions of $a^\mu_{\epsilon,\q}$ at
$x^0=-\infty$, we see that we must have
\begin{equation}
b_{\pm,\q}^\mu(x)=
\sum_{\lambda,a}\int\limits_\k
\gamma_{\pm,\q}^{\k\lambda a}\;
a_{\pm\k\lambda a}^\mu(x)\, .
\label{eq:linear1}
\end{equation}
The coefficients $\gamma_{\pm,\q}^{\k\lambda a}$ in these
linear decompositions do not depend on space or time. The boundary
conditions at $x^0=-\infty$ do not constrain further the coefficients
$\gamma_{\pm,\q}^{\k\lambda a}$, but they can be determined from the
boundary conditions at $x^0=+\infty$. To achieve this end, we 
introduce the Fourier decomposition of the functions
$a_{\pm\k\lambda a}^\mu(x)$,
\begin{equation}
a_{\pm\k\lambda a}^\mu(x)
\equiv
\sum_{\zeta,b}
\int\limits_\p
\Big\{
h_{\pm\p\zeta b}^{(+)}(x^0;\k\lambda a)\,a_{-\p\zeta b}^{0\mu}(x)
+
h_{\pm\p\zeta b}^{(-)}(x^0;\k\lambda a)\,a_{+\p\zeta b}^{0\mu}(x)
\Big\}\; .
\label{eq:small-fluct}
\end{equation}
It is then a simple exercise to rewrite the boundary conditions at
$x^0=+\infty$ as 
\begin{eqnarray}
&&
\sum_{\lambda,a}
\int\limits_\k
\Big[
\gamma_{-,\q}^{\k\lambda a}\,h_{-\k\lambda a}^{(+)}(\p\zeta b)
-
\gamma_{+,\q}^{\k\lambda a}\,h_{+\k\lambda a}^{(+)}(\p\zeta b)
\Big]
=\delta(\p-\q)\,f^{(+)}(\p\zeta b)\; ,
\nonumber\\
&&
\sum_{\lambda,a}
\int\limits_\k
\Big[
\gamma_{+,\q}^{\k\lambda a}\,h_{+\k\lambda a}^{(-)}(\p\zeta b)
-
\gamma_{-,\q}^{\k\lambda a}\,h_{-\k\lambda a}^{(-)}(\p\zeta b)
\Big]
=\delta(\p-\q)\,f^{(-)}(\p\zeta b)\; ,
\nonumber\\
&&
\label{eq:a-bc1}
\end{eqnarray}
where, to keep the expressions compact, we have omitted the argument
$x^0=+\infty$ in all the Fourier coefficients. This can be seen as a
system of linear equations for the coefficients
$\gamma_{\pm,\q}^{\k\lambda a}$. The solution of this system of linear
equations is obtained in appendix~\ref{app:Fourier}.

Inserting the results in eq.~(\ref{eq:coeff-gamma}) for
$\gamma_{\pm,q}^{\k\lambda a}$ and Fourier decomposition in
eq.~(\ref{eq:small-fluct}) in eq.~(\ref{eq:linear1}), one can easily
determine the Fourier coefficients of $b_{\pm,\q}^\mu(x)$ at
$x^0=+\infty$ (eq.~(\ref{eq:Fourier-b}). Inserting these into
eq.~(\ref{eq:ddF-LO1}), we obtain\footnote{We additionally use
  eqs.~(\ref{eq:unit}) to symmetrize the formula with respect to
  $(\p,\q)$.}
\begin{eqnarray}
&&
\left.
\frac{\delta^2 \ln {\cal F}[z]}{\delta z(\p)\delta z(\q)}
\right|_{_{\rm LO; z(\p),z(\q) = 1}}
=
\frac{1}{2}\;
\frac{1}{(2\pi)^6 4E_\p E_\q}\sum_{\lambda,a}\sum_{\xi,b}\sum_{\zeta,c}
\int\limits_\k
\nonumber\\
&&
\!\!\!\!\!
\times
\Bigg\{
\left(
h_{+\k\lambda a}^{(-)}(\p\xi b)
h_{-\k\lambda a}^{(-)}(\q\zeta c)
+
h_{-\k\lambda a}^{(-)}(\p\xi b)
h_{+\k\lambda a}^{(-)}(\q\zeta c)
\right) \;f^{(+)}(\p\xi b)f^{(+)}(\q\zeta c)
\nonumber\\
&&
+
\left(
h_{+\k\lambda a}^{(+)}(\p\xi b)
h_{-\k\lambda a}^{(+)}(\q\zeta c)
+
h_{-\k\lambda a}^{(+)}(\p\xi b)
h_{+\k\lambda a}^{(+)}(\q\zeta c)
\right) \;f^{(-)}(\p\xi b)f^{(-)}(\q\zeta c)
\nonumber\\
&&
+
\left(
h_{+\k\lambda a}^{(-)}(\p\xi b)
h_{-\k\lambda a}^{(+)}(\q\zeta c)
+
h_{-\k\lambda a}^{(-)}(\p\xi b)
h_{+\k\lambda a}^{(+)}(\q\zeta c)
\right) \;f^{(+)}(\p\xi b)f^{(-)}(\q\zeta c)
\nonumber\\
&&
\!\!+
\left(
h_{+\k\lambda a}^{(+)}(\p\xi b)
h_{-\k\lambda a}^{(-)}(\q\zeta c)
+
h_{-\k\lambda a}^{(+)}(\p\xi b)
h_{+\k\lambda a}^{(-)}(\q\zeta c)
\right) \;f^{(-)}(\p\xi b)f^{(+)}(\q\zeta c)
\Bigg\}
\nonumber\\
&&
-\frac{1}{(2\pi)^3 2E_\p}\delta(\p-\q)\sum_{\zeta,c}
 f^{(+)}(\p\zeta c)f^{(-)}(\p\zeta c)
\; .
\label{eq:d2F-LO2}
\end{eqnarray}
We have therefore obtained an expression for the connected piece of
the two gluon spectrum entirely in terms of Fourier modes of the
classical field ($f^{(\pm)}$) and the small fluctuation field ($h_{\pm
  \k \lambda a}^{(\pm)}$). The former can be determined by solving the
Yang-Mills equations with retarded boundary conditions while the
latter can be determined by solving the equations for small
fluctuations about the classical field, also with retarded boundary
conditions.

The last term in eq.~(\ref{eq:d2F-LO2}), proportional to
$\delta(\p-\q)$ times the single particle spectrum, arises because the
quantity $dN_2/d^3\p d^3\q$ is defined in such a way that its integral
over $\p$ and $\q$ gives the average value\footnote{This can easily be
  checked on a Poisson distribution, for which the second derivative
  $\delta\ln {\cal F}[z]/\delta z(\p)\delta z(\q)$ is exactly zero.
  When we insert this in eq.~(\ref{eq:2gluon-1}) and integrate over
  $\p$ and $\q$, we obtain $\langle N(N-1)\rangle=\langle
  N\rangle^2$ -- as expected for a Poisson distribution.} of $N(N-1)$.
This term provides the $-N$ contribution to this quantity. Because the
logs in the multiplicity $N$ arise only at the order ${\cal O}(g^0)$,
this term cannot provide any leading log in the 2-gluon spectrum and
can thus be dropped.

\subsection{Leading log resummation of the 2 gluon spectrum}

Combining the results in eq.~(\ref{eq:d2F-LO2}) and
eq.(\ref{eq:O-NLO1}) in eq.~(\ref{eq:N2-NLO}), we now have a formula
for the 2-gluon spectrum, including both LO and NLO contributions.
As mentioned previously, it can in principle be evaluated, in full
generality, by numerical solutions of small fluctuation partial
differential equations with retarded boundary conditions. However, if
one is interested primarily in the leading logarithmic piece of the
NLO contributions, we can go significantly further analytically.
Indeed, as we will now show by using the information obtained thus
far, we can compute the leading logarithmic contributions to the two
gluon spectrum in perturbation theory.

The first step in this derivation is to obtain an even more compact
form for eq.~(\ref{eq:d2F-LO2}) by using the linear operator
${\mathbbm T}_{\u}$ that we used previously in the expression for the
1-loop corrections to the single particle spectrum--see
eq.~(\ref{eq:O-NLO1}). In Paper I, we demonstrated explicitly that
this operator allows one to express the value of a retarded
fluctuation at a point $x$ in terms of the value of the classical
field at the same point as
\begin{equation}
a^\mu(x)=\int\limits_\Sigma d^3\vec\u\;
\big[a\cdot{\mathbbm T}_\u\big]\;{\cal A}^\mu(x)\; ,
\end{equation}
where $\Sigma$ is the initial surface on which we know the value of
the fluctuation. (The point $x$ is located above this surface.)
Performing the Fourier decomposition of both sides of this relation,
we obtain simply the relation between the Fourier coefficients (at
$x^0=+\infty$) of the small fluctuation and the classical field to be
\begin{equation}
h^{(\epsilon)}(+\infty;\p\lambda a)=\int\limits_\Sigma d^3\vec\u\;
\big[a\cdot{\mathbbm T}_\u\big]\;f^{(\epsilon)}(+\infty;\p\lambda a)\; .
\label{eq:T-fourier}
\end{equation}
Applying eq.~(\ref{eq:T-fourier}) to the various fluctuations that
appear in eq.~(\ref{eq:d2F-LO2}), and using the $z(p)=1$
simplification of eq.~(\ref{eq:dF-LO1}),
\begin{equation}
\left.\frac{dN}{d^3\p}\right|_{_{\rm LO}}
=
\left.
\frac{\delta \ln {\cal F}[z]}{\delta z(\p)}
\right|_{_{z=1,{\rm LO}}}
=
\frac{1}{(2\pi)^3 2E_\p}
\sum_{\zeta,c}
f^{(+)}(\p\zeta c)f^{(-)}(\p\zeta c)
\; ,
\end{equation}
it is a matter of simple algebra to check that 
\begin{eqnarray}
\left.
\frac{\delta^2 \ln {\cal F}[z]}{\delta z(\p)\delta z(\q)}
\right|_{_{{\rm LO};z(\p),z(\q)=1}}
\!\!\!\!\!\!\!
=
-
\delta(\p-\q)\;\left.
\frac{dN}{d^3\p}
\right|_{_{\rm LO}}
\!\!\!\!
+
\Big[{\cal L}_2
\Big]_{\rm connected}
\left.\frac{dN}{d^3\p}\right|_{_{\rm LO}}
\left.\frac{dN}{d^3\q}\right|_{_{\rm LO}}
\; .
\label{eq:N2-NLO2}
\end{eqnarray}
The subscript ``connected'' indicates that one of the ${\mathbbm T}$
operators in the expression ${\cal L}_2$ appearing in
eq.~(\ref{eq:O-NLO1}) must act on the $\p$-dependent factor and the
other on the $\q$-dependent factor. (Terms where they both act on the
same factor should be excluded.)

We see now that eqs.~(\ref{eq:N2-NLO1}) and (\ref{eq:N2-NLO2}) can be
combined very easily, because  the sum of ``disconnected'' and
``connected'' terms is equivalent to the unrestricted action of ${\mathbbm
  T}_\u{\mathbbm T}_\v$ on the product $(dN/d^3\p)(dN/d^3\q)$. We obtain thus 
\begin{eqnarray}
&&
\left.\frac{d^2 N_2}{d^3\p d^3\q}\right|_{_{\rm NLO}}
=
-\delta(\p-\q)\;\left.\frac{dN}{d^3\p}\right|_{_{\rm LO}}
+
\Big[{\cal L}_1+{\cal L}_2
\Big]\,
\left.\frac{d N}{d^3\p}\right|_{_{\rm LO}}
\left.\frac{d N}{d^3\q}\right|_{_{\rm LO}}\; ,
\nonumber\\
&&
\label{eq:N2-NLO3}
\end{eqnarray}
where ${\cal L}_1$ and ${\cal L}_2$ were both introduced previously in
eq.~(\ref{eq:O-NLO1}).

We shall now discuss the logarithmic singularities in this expression.
Firstly, $\delta(\p-\q)(dN/d^3\p)_{_{\rm LO}}$ does not contain large
logarithms in $x$ because these logs start appearing at NLO in the
single gluon spectrum. Because we are restricting our discussion to
leading logs, we can therefore discard this term henceforth.  The
logarithmic divergences in the second and third terms of the r.h.s. of
eq.~(\ref{eq:N2-NLO3}) can be extracted straightforwardly by using the
main result of Paper I,
\begin{eqnarray}
{\cal L}_1+{\cal L}_2
\empile{=}\over{\rm LLog}
\ln\left(\frac{\Lambda^+}{M^+}\right)
{\cal H}_1 
+
\ln\left(\frac{\Lambda^-}{M^-}\right)
{\cal H}_2\; .
\label{eq:JIMWLK}
\end{eqnarray}
Here ${\cal H}_{1,2}$ are the JIMWLK Hamiltonians of the nuclei moving
in the $+z$ and $-z$ directions
respectively~\cite{GelisLV3,McLer1,IancuLM3,IancuV1}, $\Lambda^\pm$
represent the longitudinal momenta that separate the static color
sources $\rho_{1,2}$ in each of the nuclei respectively from the the
gauge fields that produce gluons at the rapidity of interest, and
$M^\pm$ corresponds to the typical longitudinal momentum scales of the
object (the two gluon spectrum in this case) to which the operator is
applied. From eq.~(\ref{eq:JIMWLK}) we obtain
\begin{equation}
\left.\frac{d^2 N_2}{d^3\p d^3\q}\right|_{_{\rm LO + NLO}} 
\empile{=}\over{\rm LLog} 
\Bigg[1+\ln\left(\frac{\Lambda^+}{M^+}\right)
{\cal H}_1 + \ln\left(\frac{\Lambda^-}{M^-}\right)
{\cal H}_2\Bigg]\,
\left.\frac{dN}{d^3 p}\right|_{_{\rm LO}}\; 
\left.\frac{dN}{d^3 p}\right|_{_{\rm LO}}\; .
\label{eq:fact-formula2}
\end{equation}
All of our discussion thus far has been for a fixed distribution of
sources $\rho_{1,2}$ in the two nuclei. The CGC effective theory
~\cite{McLerV1,McLerV2,McLerV3,JalilKMW1,JalilKLW1,JalilKLW2,JalilKLW3,JalilKLW4,IancuLM1,IancuLM2,FerreILM1},
prescribes to average physical quantities over all the possible
configurations $\rho_{1,2}$ of the fast color sources representing the
projectiles, with gauge invariant weight functionals $W[\rho_{1,2}]$
that describe the probability of each configuration. When we integrate
eq.~(\ref{eq:fact-formula2}) over $\rho_{1,2}$, we can exploit the
hermiticity of the JIMWLK Hamiltonians ${\cal H}_{1,2}$ in order to
integrate by parts, so that the Hamiltonians are now acting on the
distributions $W[\rho_{1,2}]$. By reproducing the arguments developed
in Paper I for the single gluon spectrum, we obtain finally the
factorization formula for inclusive two-gluon production,
\begin{equation}
\left<\frac{d^2 N_2}{d^3\p d^3\q}\right>_{_{\rm LLog}}
=
\int \big[D\rho_1\big]\big[D\rho_2\big]\;
W_{Y_1}\big[\rho_1\big]\,
W_{Y_2}\big[\rho_2\big]\;
\left.\frac{dN}{d^3\p}\right|_{_{\rm LO}}
\left.\frac{dN}{d^3\q}\right|_{_{\rm LO}}\; ,
\label{eq:2gluon-fact}
\end{equation}
at leading log accuracy. 
Here the distributions $W[\rho_{1,2}]$ obey the JIMWLK equation 
\begin{equation}
\frac{\partial W_{_Y}[\rho]}{\partial Y} = {\cal H}\; W_{_Y}[\rho] \; ,
\label{eq:H-JIMWLK1}
\end{equation}
and are evolved thus from non-perturbative initial conditions at the
beam rapidities to the rapidities $Y_1=\ln(\sqrt{s}/M^+)$ and
$Y_2=\ln(\sqrt{s}/M^-)$ respectively. In the regime where gluon
radiation between the two tagged gluons is small, this formula resums
all leading logarithms of $1/x_{1,2}$ as well as all the rescattering
corrections in $(g\rho_{1,2})^n$ to all orders.

We now address the primary limitation of the present calculation. As
the previous discussion hints, it is valid when the momenta $\p$ and
$\q$ of the two observed gluons are close enough in rapidity so that
they have similar longitudinal components. More precisely, we need to
have
\begin{eqnarray}
\alpha_s \ln\left(\frac{p^+}{q^+}\right)\ll 1\quad,\qquad
\alpha_s \ln\left(\frac{p^-}{q^-}\right)\ll 1\; .
\label{eq:condition}
\end{eqnarray}
If this is the case, we can simply take $M^\pm$ to be the common
value\footnote{It is of course not necessary that $p^+$ and $q^+$ be
  equal, just that they are close enough so that it does not matter
  which value we chose between $p^+$ and $q^+$.} of $p^\pm,q^\pm$.
Physically, the condition of eq.~(\ref{eq:condition}) means that the
probability of radiating a gluon between the two measured gluons is
small. When the rapidity separation between the two gluons is large
such that eq.~(\ref{eq:condition}) is violated, we need to resum gluon
emissions between the tagged gluons; this would require a
generalization of the present formalism, which is not discussed here.

\subsection{Factorization and the ridge in AA collisions}
\label{sec:ridge}

A striking ``ridge'' structure has been revealed in studies of the
near side spectrum of correlated pairs of hadrons by the STAR
collaboration~\cite{Adamsa4,Wang1,Adamsa5}. The spectrum of correlated
pairs on the near side of the detector (defined by an accompanying
unquenched jet spectrum) extends across the entire detector acceptance
in pseudo-rapidity of order $\Delta \eta\sim 2$ units but is strongly
collimated for azimuthal angles $\Delta \phi$.  Preliminary analyses
of measurements by the PHENIX~\cite{Adarea1} and PHOBOS~\cite{Wosie1}
collaborations appear to corroborate the STAR results.  In the latter
case, with a high momentum trigger, the ridge is observed to span the
wider PHOBOS acceptance in pseudo-rapidity of $\Delta \eta \sim 4$
units.

In Ref.~\cite{DumitGMV1}, it was argued that the ridge is formed as a
consequence of both long range rapidity correlations that are generic
in hadronic and nuclear collisions at high energies, plus the radial
flow of the hot partonic matter that is specific to high energy
nuclear collisions.  Let us first focus on the long range correlations
that are essential to this picture--how are they generated?

In the leading order formalism of the CGC, classical solutions of
Yang-Mills equations are boost
invariant~\cite{KovneMW1,KovneMW2,KovchR1,GyulaM1}.  Real time
numerical
simulations~\cite{KrasnV1,KrasnV2,KrasnV3,KrasnNV1,KrasnNV2,KrasnNV3,Lappi1,Lappi2,Lappi3}
also demonstrate that the Yang-Mills fields form flux tubes of a
typical transverse size $1/Q_s$ (where $Q_s$ is the saturation scale)
with parallel chromo--electric and chromo--magnetic field strengths.
(An important consequence is that these Glasma fields~\cite{LappiM1}
have non-trivial topological charge~\cite{KharzKV1}.) Now, in
section~\ref{subsec:2gluon-LO}, we showed that the leading order
2-gluon spectrum, for a fixed configuration of sources, was given by
eq.~(\ref{eq:N2-LO}). Because each of the single particle
distributions is boost invariant, the two particle spectrum is also,
at this order, independent of the rapidity separation of the gluons.
While the two gluons are uncorrelated for a fixed configuration of
sources, correlations are built in through the averaging over the
source distributions. In Ref.~\cite{DumitGMV1}, the source
distribution was assumed to be Gaussian as in the
McLerran--Venugopalan (MV) model~\cite{McLerV1,McLerV2,McLerV3}. The
ridge spectrum was shown to have the simple form
\begin{equation}
\frac{\Delta \rho}{\sqrt{\rho_{\rm ref}}}
\equiv C(\p,\q)\,
\frac{\left< \frac{dN}{dy} \right>}
{
\left<\frac{dN}{dy_p\, p dp \,d\phi_p}\right> 
\left<\frac{dN}{dy_q \,q dq \,d\phi_q}\right>
}
= \frac{K_{_N}}{\alpha_s (Q_s)} \; ,
\label{eq:Glasma-tube1}
\end{equation}
where 
\begin{eqnarray}
C(\p,\q) 
\equiv
\left<\frac{dN_2}{dy_p d^2\p_\perp dy_q d^2\q_\perp}\right> 
 - 
\left<\frac{dN}{dy_p d^2\p_\perp}\right> 
\left<\frac{dN}{dy_q d^2\q_\perp}\right>\; , 
\label{eq:C-def}
\end{eqnarray}
and $K_N$ is a number of order unity. For further details, we refer
the reader to Ref.~\cite{DumitGMV1}.

There are several conceptual issues in this context. Firstly, how does
one justify this averaging procedure for the 2-gluon spectrum from
first principles? Secondly, how does one build in energy evolution of
the sources? And finally, do NLO contributions spoil this picture?
Our results in this paper solve most of these conceptual issues. Our
result, in eq.~(\ref{eq:2gluon-fact}), shows that the trivial LO
result of eq.~(\ref{eq:N2-LO}) can be promoted to a full Leading Log
result simply by averaging it over the sources $\rho_{1,2}$--with
distributions of sources that evolve according to the JIMWLK equation.
Most importantly, this shows that all higher order corrections, to
leading logs in $x_{1,2}$, do not spoil the form in
eq.~(\ref{eq:Glasma-tube1}) of the Glasma flux tube picture and
provides compelling evidence that it is a robust result beyond LO. As
discussed previously, this picture will have to be modified when the
rapidity separation between the gluons is greater than $\alpha_s^{-1}(Q_s)$.

These initial state considerations are not affected by the final state
transverse flow of the Glasma flux tubes which is the other important
feature determining the near side ridge seen in heavy ion collisions.
Its been shown very recently that a proper treatment of flow and
hadronization effects of the Glasma flux tubes provides excellent
quantitative agreement with the RHIC data on the dependence of the
ridge amplitude on centrality and as a function of energy, as well as
the angular width of the ridge as a function of
centrality~\cite{GavinMM1}. Further sophisticated treatments of both
the initial state effects discussed here and the final state effects
discussed in Ref.~\cite{GavinMM1}, therefore open the door to
quantitative 3-D imaging of heavy ion collisions.  A deeper relation
between initial and final state effects, as outlined in Paper I, can
be obtained by studying quantum fluctuations at NLO, that are not
accompanied by logs in $x_{1,2}$, but grow rapidly in
time~\cite{RomatV1,RomatV2} in a manner analogous to plasma
instabilities~\cite{Mrowc1}.

We should also mention that the initial state effects described here
are also present in proton/deuteron-nucleus
collisions~\cite{BaierKNW1,JalilK2,FukusH1}, without the final state
effects characteristic of the ridge in nucleus-nucleus collisions.
These collisions are therefore useful in order to isolate the initial
state effects and to corroborate the framework of multiparticle
production in high energy QCD developed here.

\section{Multigluon inclusive spectrum}
\label{sec:multi}

In this section, we will show how the results of the previous section
modify multi-gluon probability distributions, with the
caveat, as previously, that these gluons are emitted in a narrow
rapidity window. We will also derive a simple expression for the
differential probability of producing $n$ gluons.

\subsection{$n$-gluon spectrum at LO and NLO}
Our starting point in evaluating the inclusive $n$-gluon spectrum is
eq.~(\ref{eq:ngluon}). Because we have thus far obtained
expressions up to NLO for the first and second derivatives of
$\ln{\cal F}[z]$, it is convenient to rewrite this expression
as\footnote{This formula is obtained by replacing ${\cal F}[z]$ by
  $\exp(\ln{\cal F}[z])$ in eq.~(\ref{eq:ngluon}).}
\begin{eqnarray} 
\underbrace{
\frac{d^n N_n}{d^3\p_1\cdots d^3\p_n}
}_{\scriptstyle{{\cal O}\left(\frac{1}{g2n}+\cdots\right)}}
&=&
\underbrace{
\prod_{i=1}^n\frac{\delta \ln{\cal F}[z]}{\delta z(\p_i)}
}_{\scriptstyle{{\cal O}\left(\frac{1}{g2n}+\cdots\right)}}
\nonumber\\
&+&
\underbrace{
\sum_{i<j}\frac{\delta^2 \ln{\cal F}[z]}{\delta z(\p_i)\delta z(\p_j)}
\prod_{k\not=i,j}\frac{\delta\ln{\cal F}[z]}{\delta z(\p_k)}
}_{\scriptstyle{{\cal O}\left(\frac{1}{g2(n-1)}+\cdots\right)}}
+\cdots
\label{eq:moments}
\end{eqnarray}
Because $\ln{\cal F}[z]={\cal O}(g^{-2})$ in our power counting, the
LO term in the r.h.s. is of order $g^{-2n}$, the NLO term is of order
$g^{-2(n-1)}$, and NNLO and higher terms represented by the ellipses
are omitted at the level of the present discussion.
The $n$-gluon spectra on the l.h.s. of eq.~(\ref{eq:moments}) 
are quantities that, for $n>1$, are given by the first term
on the r.h.s. By computing them to NLO we
gain access to the first correction to the Poisson distribution,
the deviation of the variance of the multiplicity distribution from the
Poissonian result $\langle N(N-1)\rangle = \langle N \rangle^2$ 
and the corresponding modifications for the higher moments of the 
distribution. We refer to the appendix \ref{app:poisson} for a more
detailed discussion of the interpretation of our result for the probability
distribution of the gluon multiplicity.

At leading order, only the first term contributes, and we obtain (for
a fixed distribution of sources)
\begin{equation}
\left.
\frac{d^n N_n}{d^3\p_1\cdots d^3\p_n}
\right|_{_{\rm LO}}
=
\prod_{i=1}^n\left.\frac{dN}{d^3\p_i}\right|_{_{\rm LO}}\; .
\end{equation}
At next to leading order, we have
\begin{eqnarray}
&&
\left.
\frac{d^n N_n}{d^3\p_1\cdots d^3\p_n}
\right|_{_{\rm NLO}}
=
\sum_{i=1}^n
\left.\frac{dN}{d^3\p_i}\right|_{_{\rm NLO}}
\prod_{j\not=i}
\left.\frac{dN}{d^3\p_j}\right|_{_{\rm LO}}
\nonumber\\
&&\qquad\qquad\qquad\qquad\qquad\qquad\qquad
+
\sum_{i<j}
\left.
\frac{\delta^2 \ln{\cal F}[z]}{\delta z(\p_i)\delta z(\p_j)}\right|_{_{\rm LO}}
\prod_{k\not=i,j}\left.\frac{dN}{d^3\p_k}\right|_{_{\rm LO}}\; .
\end{eqnarray}
All the objects that appear in this equation are known already from
the discussion in Paper I and the previous section. In
\cite{GelisLV3}, we showed that
\begin{equation}
\left.\frac{dN}{d^3\p}\right|_{_{\rm NLO}}
= \Big[{\cal L}_1+{\cal L}_2\Big]\; \left.\frac{dN}{d^3\p}\right|_{_{\rm LO}}
+\Delta N_{_{\rm NLO}}(\p)
\; ,
\label{eq:NLO1}
\end{equation}
where ${\cal L}_1$ and ${\cal L}_2$ are defined in
eq.~(\ref{eq:O-NLO1}). In the previous section, we showed
that\footnote{We are ignoring the term
  $-\delta(\p-\q)\left.\frac{dN}{d^3\p}\right|_{_{\rm LO}}$ because it
  does not contribute in the leading logarithmic approximation in $x$
  as discussed previously.}
\begin{equation}
\left.
\frac{\delta^2 \ln{\cal F}[z]}{\delta z(\p)\delta z(\q)}\right|_{_{\rm LO}}
=
\Big[{\cal L}_2\Big]_{\rm connected} 
\left.\frac{dN}{d^3\p}\right|_{_{\rm LO}}
\left.\frac{dN}{d^3\q}\right|_{_{\rm LO}}\; ,
\label{eq:NLO2}
\end{equation}
where we remind the reader that the subscript ``connected'' attached
to the operator ${\cal L}_2$ indicates that the two operators
${\mathbbm T}$ it contains do not simultaneously act on the same
object.

\subsection{Leading Log resummation}
If we combine the terms in 
eqs.~(\ref{eq:NLO1}) and (\ref{eq:NLO2}), we get simply
\begin{equation}
\left.
\frac{d^n N_n}{d^3\p_1\cdots d^3\p_n}
\right|_{_{\rm NLO}}
\empile{=}\over{\rm LLog}
\Big[{\cal L}_1+{\cal L}_2\Big]\,
\prod_{i=1}^n
\left.\frac{dN}{d^3\p_i}\right|_{_{\rm LO}}\; .
\end{equation}

Using again eq.~(\ref{eq:JIMWLK}) and following the steps that lead
from eq.~(\ref{eq:fact-formula2}) to eq.~(\ref{eq:2gluon-fact}), we
arrive at the all order leading log $n$-gluon spectrum
\begin{eqnarray}
  &&
  \left<\frac{d^n  N_n}{d^3\p_1\cdots d^3\p_n}\right>_{_{\rm LLog}}
  =
  \int \big[D\rho_1\big]\big[D\rho_2\big]\;
  W_{Y_1}\big[\rho_1\big]\,
  W_{Y_2}\big[\rho_2\big]\;
  \nonumber\\
  &&\qquad\qquad\qquad\qquad\qquad\qquad\qquad\qquad\qquad
  \times
  \left.\frac{dN}{d^3\p_1}\right|_{_{\rm LO}}\cdots
  \left.\frac{dN}{d^3\p_n}\right|_{_{\rm LO}}\; .
\label{eq:ngluon-LLog}
\end{eqnarray}
Once again, one needs all the rapidity differences between the $n$
measured gluons to be much smaller than $\alpha_s^{-1}$, to ensure all 
leading logarithmic contributions are resummed by this formula.

\subsection{Generating functional in a small rapidity slice}
Eq.~(\ref{eq:ngluon-LLog}) provides a complete description of gluon
production in the leading log $x$ approximation when one considers a
slice in rapidity of width $\Delta Y\ll \alpha_s^{-1}$. One can
summarize these results into a generating functional ${\cal
  F}_{_{Y,\Delta Y}}[z(\p)]$ defined from the ``master'' ${\cal
  F}[z(\p)]$ as
\begin{equation}
{\cal F}_{_{Y,\Delta Y}}[z(\p)]
=
{\cal F}[z^*(\p)]
\quad\mbox{with}\;\;\left\{
\begin{aligned}
z^*(\p)&=&
z(\p)&&\quad\mbox{if\ \ }y_p\in
\scriptstyle{\big[Y-\frac{\Delta Y}{2},Y+\frac{\Delta Y}{2}\big]}&&&\\
z^*(\p)&=&1&&\quad\mbox{otherwise}&&&
\end{aligned}
\right.\; .
\end{equation}
Setting the argument of the generating functional to unity outside of
the phase space region of interest means that we define observables
that are completely inclusive with respect to this unobserved part of
the phase space.

We see from eq.~(\ref{eq:rep1}) that ${\cal F}[z^*(\p)]$ can be
obtained by multiplying eq.~(\ref{eq:ngluon-LLog}) by
$(z^*(\p_1)-1)\cdots (z^*(\p_n)-1)/n!$, integrating over the $n$-gluon
phase space and summing over $n$.  Because $z^*(\p)$ is unity outside
of the strip of width $\Delta Y$ in rapidity, the $n$-gluon spectrum
outside of the regime of validity of eq.~(\ref{eq:ngluon-LLog}) is not
needed. This procedure leads to a simple exponentiation of the leading
log factorized formula for the generating functional ${\cal
  F}_{_{Y,\Delta Y}}$ as
\begin{eqnarray}
&&
\left<{\cal F}_{_{Y,\Delta Y}}[z(\p)]\right>_{_{\rm LLog}}
=
\int \big[D\rho_1\big]\big[D\rho_2\big]\;
W_{Y_{\rm beam}-Y}\big[\rho_1\big]\,
W_{Y_{\rm beam}+Y}\big[\rho_2\big]\;
\nonumber\\
&&\qquad\qquad\qquad\qquad\qquad
\times\,
\exp\left[\;\;\int\limits_{Y-\frac{\Delta Y}{2}}^{Y+\frac{\Delta Y}{2}} d^3\p\;
(z(\p)-1)\;\left.\frac{dN}{d^3\p}\right|_{_{\rm LO}}\right]
\; .
\label{eq:Fz-slice}
\end{eqnarray}
This leading log result for the generating functional, in turn, allows us to 
extract the corresponding formula for the differential probability of
producing exactly $n$ gluons in the rapidity slice of interest. This gives   
\begin{eqnarray}
&&\left<\frac{d^n P_n}{d^3\p_1\cdots d^3\p_n}\right>_{_{\rm LLog}}
=
\int \big[D\rho_1\big]\big[D\rho_2\big]\;
  W_{Y_{\rm beam}-Y}\big[\rho_1\big]\,
  W_{Y_{\rm beam}+Y}\big[\rho_2\big]\nonumber\\
  &&\qquad\qquad
  \times\,
  \frac{1}{n!}
  \left.\frac{dN}{d^3\p_1}\right|_{_{\rm LO}}\cdots
  \left.\frac{dN}{d^3\p_n}\right|_{_{\rm LO}}\;
  \exp\left[-\;\int\limits_{Y-\frac{\Delta Y}{2}}^{Y+\frac{\Delta Y}{2}} d^3\p\;\left.\frac{dN}{d^3\p}\right|_{_{\rm LO}}\right]
\; .
\label{eq:Pn-final}
\end{eqnarray}
This simple result, valid, we emphasize, in the leading log
approximation, suggests that the particle distribution in a small
rapidity slice can be written as the average over $\rho_{1,2}$ of a
Poisson distribution with the leading log corrections
completely factorized into the JIMWLK evolution of the sources.
Note that, despite appearances,  eq.~(\ref{eq:Pn-final}) is
not a Poisson distribution after the integration over the sources,
because particles
produced uncorrelated in each configuration of $\rho_1$ and $\rho_2$
are correlated in the averaged distribution because of the
correlations among the color sources\footnote{For instance, two color
  sources may be correlated because they result from the splitting of
  a common ``ancestor'' in the course of JIMWLK evolution.}.

In general, even for a fixed distribution of sources, the probability
distribution is not Poissonian~\cite{GelisV2}. To some extent, the
fact that we get a Poissonian functional form in the integrand of
eq.~(\ref{eq:Pn-final}) is a consequence of the way we have organized
our calculation.  In eq.~(\ref{eq:moments}) we are performing a weak
coupling expansion of the moments $\langle N(N-1)\cdot \dots \cdot
(N-n+1)\rangle$, that includes the orders $g^{-2n}$ and the leading
log part of the order $g^{-2(n-1)}$. Terms starting at the order
$g^{-2(n-2)}$ are beyond the accuracy of our calculation, and
therefore their value in our formulas are arbitrary.  The
arbitrariness of these subleading terms influences the precise form of
the resulting generating functional.  For example, if we had performed
the weak coupling expansion of $\langle N^n\rangle$ instead of
$\langle N(N-1)\cdot \dots \cdot (N-n+1)\rangle$, we would have
obtained a different generating functional. Of course, the two
generating functionals so obtained would lead to the same moments of
the distribution to the order of our calculation.
The nontrivial aspect of our result in eq.~(\ref{eq:Pn-final}) is that
all the deviations from a Poisson distribution that result from the
large logarithms of $x$ at NLO can be factorized into the JIMWLK
evolution of the sources.  Equation~(\ref{eq:Pn-final}) shows how these
corrections modify the $n$ gluon production probabilities.  The
Poissonian nature of the multiplicity distribution and deviations from
it are discussed in more detail in appendix \ref{app:poisson}.

\section{Conclusion and outlook}
We demonstrated in this paper that our result of Paper I on initial
state JIMWLK factorization for the single inclusive gluon spectrum in
nucleus-nucleus collisions can be extended to inclusive multigluon
spectra. Our result is valid provided all the gluons are produced in a
rapidity window of width $\Delta Y \lesssim \alpha_s^{-1}$.  Our final
result for the generating functional for multigluon production, in the
leading logarithmic approximation in $x_{1,2}$, is very simple; the
distribution of gluons produced in the stated rapidity window can be
written as the average over the JIMWLK-evolved distributions of
sources of a Poisson distribution.  It is important to keep in mind
that the result of this source average is not a Poisson distribution,
due to the correlations between the evolved color sources.

As we discussed in section~\ref{sec:ridge}, our results are of great
interest in detailed imaging of the space--time evolution of
nucleus--nucleus collisions. An important ingredient in future studies
will be to extend the present result to the case of correlations
between gluons produced at rapidity differences
$\alpha_s^{-1}\lesssim\Delta Y$.
A full leading log computation of these initial long range rapidity
correlations requires that one identifies and resums the additional
large logarithmic corrections that may arise when the rapidities in
the two-gluon spectrum are widely separated.

An important caveat (also applicable to our previous study of the
single gluon spectrum in nucleus-nucleus collisions) is that final
state effects, related to the growth of unstable fluctuations, need to
be resummed. While the details are still unknown, the structure of the
result is known. The result of the resummation of unstable
fluctuations, as shown in Paper I, can be expressed as
\begin{eqnarray}
&&
\left<{\cal O}\right>_{\rm LLog+LInst}
=
\int
\big[D\wt{\cal A}^+_1\big]\big[D\wt{\cal A}^-_2\big]\;
W_{_{Y_1}}\big[\wt{\cal A}^+_1\big]\,
W_{_{Y_2}}\big[\wt{\cal A}^-_2\big]\,
\nonumber\\
&&\qquad\qquad\qquad\times
\int\big[Da(\vec\u)\big]\;\widetilde{Z}[a(\vec\u)]\;
{\cal O}_{_{\rm LO}}[\wt{\cal A}^+_1+a,\wt{\cal A}^-_2+a]
\; .
\label{eq:final}
\end{eqnarray}
Here, we have traded the sources $\tilde\rho_{1,2}$ in covariant gauge
for the corresponding gauge fields $\wt{\cal A}^\pm_{1,2} \equiv
\frac{1}{\nabla_\perp^2}{\tilde \rho}_{1,2}$. The functional
$\widetilde{Z}[a(\vec\u)]$ is the spectrum of small fluctuations of
the classical field on the forward light-cone. In Paper I, ${\cal O}$
corresponded to the single inclusive spectrum but this formula also
applies to the multigluon spectrum because the proof does not depend
on the nature of the observable being measured.  However, the complete
functional form of $\widetilde{Z}[a(\vec\u)]$ is still unknown--for a
first attempt, see Ref.~\cite{FukusGM1}.

These considerations are eased somewhat if we take the
``dilute--dense'' limit of proton/deuteron--nucleus collisions because
we don't expect instabilities to play a major role in that case.
Several studies have been performed in this
limit~\cite{BaierKNW1,JalilK2,FukusH1,Braun5,Marqu1,KovneL1}. A
particular focus is on the applicability of the so called AGK cutting
rules~\cite{AbramGK1,CiafaM1,BarteSV1,BarteSV2}. We plan to address
these issues in a future work.

\section*{Acknowledgements}
R. V.'s work is supported by the
US Department of Energy under DOE Contract No.  DE-AC02-98CH10886.
F.G.'s work is supported in part by Agence Nationale de la Recherche
via the programme ANR-06-BLAN-0285-01.

\appendix

\section{Fourier coefficients of small fluctuation fields}
\label{app:Fourier}

We will outline here the solution to the system of equations 
\begin{eqnarray}
&&
\sum_{\lambda,a}
\int\limits_\k
\Big[
\gamma_{-,\q}^{\k\lambda a}\,h_{-\k\lambda a}^{(+)}(\p\zeta b)
-
\gamma_{+,\q}^{\k\lambda a}\,h_{+\k\lambda a}^{(+)}(\p\zeta b)
\Big]
=\delta(\p-\q)\,f^{(+)}(\p\zeta b)\; ,
\nonumber\\
&&
\sum_{\lambda,a}
\int\limits_\k
\Big[
\gamma_{+,\q}^{\k\lambda a}\,h_{+\k\lambda a}^{(-)}(\p\zeta b)
-
\gamma_{-,\q}^{\k\lambda a}\,h_{-\k\lambda a}^{(-)}(\p\zeta b)
\Big]
=\delta(\p-\q)\,f^{(-)}(\p\zeta b)\; ,
\nonumber\\
&&
\end{eqnarray}
that was obtained in eq.~(\ref{eq:a-bc1}). 
We had previously derived analogous equations in the case of a simpler
scalar theory in \cite{GelisV2}.  However, in \cite{GelisV2}, we did
not manage to solve these equations and suggested that one may have to
solve them numerically. It turns out that one can in fact obtain an
analytical solution of the eqs.~(\ref{eq:a-bc1}), thanks to the relations
\begin{eqnarray}
&&
\smash{
\sum_{\lambda,a}
\int\limits_\k
}
\Big[
h_{-\k\lambda a}^{(+)}(\p\xi b)
h_{+\k\lambda a}^{(-)}(\q\zeta c)
\nonumber\\
&&\qquad\qquad\qquad
-
h_{+\k\lambda a}^{(+)}(\p\xi b)
h_{-\k\lambda a}^{(-)}(\q\zeta c)
\Big]
=
(2\pi)^3 \delta_{\xi \zeta}\,\delta_{bc}\,2E_\p\delta(\p-\q)\; ,
\nonumber\\
&&
\smash{
\sum_{\lambda,a}
\int\limits_\k
}
\Big[
h_{+\k\lambda a}^{(-)}(\p\xi b)
h_{-\k\lambda a}^{(+)}(\q\zeta c)
\nonumber\\
&&\qquad\qquad\qquad
-
h_{-\k\lambda a}^{(-)}(\p\xi b)
h_{+\k\lambda a}^{(+)}(\q\zeta c)
\Big]
=
(2\pi)^3 \delta_{\xi \zeta}\,\delta_{bc}\,2E_\p\delta(\p-\q)\; ,
\nonumber\\
&&
\sum_{\lambda,a}
\int\limits_\k
\Big[
h_{+\k\lambda a}^{(+)}(\p\xi b)
h_{-\k\lambda a}^{(+)}(\q\zeta c)
-
h_{-\k\lambda a}^{(+)}(\p\xi b)
h_{+\k\lambda a}^{(+)}(\q\zeta c)
\Big]
=0\; ,
\nonumber\\
&&
\sum_{\lambda,a}
\int\limits_\k
\Big[
h_{+\k\lambda a}^{(-)}(\p\xi b)
h_{-\k\lambda a}^{(-)}(\q\zeta c)
-
h_{-\k\lambda a}^{(-)}(\p\xi b)
h_{+\k\lambda a}^{(-)}(\q\zeta c)
\Big]
=0
\; .
\label{eq:unit}
\end{eqnarray}
These relations are the mathematical consequence of the unitary
temporal evolution of small fluctuations on top of the classical field
${\cal A}(x)$. In particular, an orthonormal basis of solutions of
eq.~(\ref{eq:fluct}) remains orthonormal at any later time. A proof of
these formulas is presented in appendix \ref{app:unit}.  Thanks to
these relations, it is easy to invert the system of equations
(\ref{eq:a-bc1}), and one gets
\begin{eqnarray}
&&
\gamma_{+,\q}^{\k\lambda a}
=
\frac{1}{(2\pi)^3 2E_\q}
\sum_{\zeta,b}
\Big[
h_{-\k\lambda a}^{(-)}(\q\zeta b)\,f^{(+)}(\q\zeta b)
+
h_{-\k\lambda a}^{(+)}(\q\zeta b)\,f^{(-)}(\q\zeta b)
\Big]
\; ,
\nonumber\\
&&
\gamma_{-,\q}^{\k\lambda a}
=
\frac{1}{(2\pi)^3 2E_\q}
\sum_{\zeta,b}
\Big[
h_{+\k\lambda a}^{(-)}(\q\zeta b)\,f^{(+)}(\q\zeta b)
+
h_{+\k\lambda a}^{(+)}(\q\zeta b)\,f^{(-)}(\q\zeta b)
\Big]\; .
\nonumber\\
&&
\label{eq:coeff-gamma}
\end{eqnarray}

\section{Unitary evolution of small fluctuations}
\label{app:unit}
Consider the partial differential equation
\begin{equation}
\Big[
(\square_x g_{\mu}{}^{\nu}-\partial_{x\mu}\partial_x^\nu)\delta^{ab}
-
\frac{\partial U({\cal A}_\epsilon)}{\partial {\cal A}_{\epsilon a\nu}(x)\partial {\cal A}_{\epsilon b}^\mu(x)}
\Big]\,a^{\mu b}(x)
=0\; ,
\label{eq:PDE}
\end{equation}
where we have written explicitly all the color indices.  We assume
that the background color field in which the wave propagates is real.
For a generic solution $a(x)$ of this equation, define the following
vectors~:
\begin{equation}
\big|{\bs a}\big>\equiv
\begin{pmatrix}
a^{\mu a}(x)\\\dot{a}^{\mu a}(x)\\
\end{pmatrix}
\quad,\quad
\big<{\bs a}\big|\equiv
\begin{pmatrix}
a^{*\mu a}(x)&\dot{a}^{*\mu a}(x)\\
\end{pmatrix}
\; ,
\label{eq:bra-ket-def}
\end{equation}
where the dot means a derivative with respect to time.  Then, it is
trivial to check that the following ``scalar product'',
\begin{equation}
\big<{\bs a}_1\big|{\bs\sigma}_2\big|{\bs a}_2\big>\equiv 
i\,g_{\mu\nu}\delta_{ab}\int d^3\x\;\Big[
\dot{a}_1^{*\mu a}(x)a_2^{\nu b}(x)
-
a_1^{*\mu a}(x)\dot{a}_{2}^{\nu b}(x)
\Big]\; ,
\label{eq:PDE-scal}
\end{equation}
where ${\bs\sigma}_2$ is the second Pauli matrix, is independent of
time when $a_1^\mu$ and $a_2^\mu$ are two solutions of
eq.~(\ref{eq:PDE}). 

Then, if the $a_{\pm\k\lambda a}(x)$ are the retarded solutions of
eq.~(\ref{eq:PDE}) whose initial conditions at $x^0\to-\infty$ are
$\epsilon^\mu_\lambda(\k) T^a e^{\pm ik\cdot x}$, one can check
explicitly that
\begin{eqnarray}
&&
\big<{\bs a}_{+\k\lambda a}\big|{\bs\sigma}_2
\big|{\bs a}_{+\k\prime\lambda^\prime a^\prime}\big>
=
(2\pi)^3 2E_\k
\,\delta_{\lambda\lambda^\prime}\,\delta_{aa^\prime}\,
\delta(\k-\k^\prime)\; ,
\nonumber\\
&&
\big<{\bs a}_{-\k\lambda a}\big|{\bs\sigma}_2
\big|{\bs a}_{-\k\prime\lambda^\prime a^\prime}\big>
=
-(2\pi)^3 2E_\k
\,\delta_{\lambda\lambda^\prime}\,\delta_{aa^\prime}\, \delta(\k-\k^\prime)\; ,
\nonumber\\
&&
\big<{\bs a}_{+\k\lambda a}\big|{\bs\sigma}_2
\big|{\bs a}_{-\k\prime\lambda^\prime a^\prime}\big>
=
\big<{\bs a}_{-\k\lambda a}\big|{\bs\sigma}_2
\big|{\bs a}_{+\k\prime\lambda^\prime a^\prime}\big>
=0\; .
\end{eqnarray}
(Since all these scalar products are time independent, it is
sufficient to check these relations by calculating the integral in the
r.h.s. of eq.~(\ref{eq:PDE-scal}) for the corresponding initial
conditions.)

Consider now a generic solution $a^\mu(x)$ of eq.~(\ref{eq:PDE}).
Since the solutions $a_{\pm\k\lambda a}^\mu(x)$ span the entire space
of solutions, we can write
\begin{equation}
\big|{\bs a}\big>
\equiv
\sum_{\lambda, a}
\int\limits_\k\;
\Big[
\gamma^{\k\lambda a}_-\,\big|{\bs a}_{-\k\lambda a}\big>
+
\gamma^{\k\lambda a}_+\,\big|{\bs a}_{+\k\lambda a}\big>
\Big]\; ,
\label{eq:decomp}
\end{equation}
where the coefficients $\gamma^{\k\lambda a}_\pm$ do not depend on
time. By using the orthogonality relations obeyed by the vectors
$\big|{\bs a}_{\pm\k\lambda a}\big>$, one obtains
\begin{equation}
\gamma^{\k\lambda a}_-
=
-\big<{\bs a}_{-\k\lambda a}\big|{\bs\sigma}_2\big|{\bs a}\big>
\quad,\quad
\gamma^{\k\lambda a}_+
=
\big<{\bs a}_{+\k\lambda a}\big|{\bs\sigma}_2\big|{\bs a}\big>
\; .
\label{eq:coeffs}
\end{equation}
Inserting these relations back into eq.~(\ref{eq:decomp}), one gets
the following identity,
\begin{equation}
\sum_{\lambda, a}
\int\limits_\k\;
\Big[
\big|{\bs a}_{+\k\lambda a}\big>\big<{\bs a}_{+\k\lambda a}\big|
-
\big|{\bs a}_{-\k\lambda a}\big>\big<{\bs a}_{-\k\lambda a}\big|
\Big]
={\bs\sigma}_2g^{\mu\nu}\delta^{bc}\; ,
\label{eq:projector}
\end{equation}
which is valid over the space of solutions of eq.~(\ref{eq:PDE}).
(The Lorentz indices $\mu,\nu$ and color indices $b,c$ do not appear
explicitly in the l.h.s., but are part of the definition of the
vectors $\big|{\bs a}\big>$ and $\big<{\bs a}\big|$--see
eq.~(\ref{eq:bra-ket-def}).) This relation is valid at all times, and
is the expression of the fact that the unitary evolution of small
fluctuations preserves the completeness of the set of states
$\big|{\bs a}_{\pm\k\lambda a}\big>$

Let us now introduce states $\big|{\bs a}_{\pm\k\lambda a}^0\big>$,
that are the analogue of the states $\big|{\bs a}_{\pm\k\lambda
  a}\big>$ in the vacuum (i.e. when the background field is zero).
Naturally, they are just plane waves $a^{0\mu}_{\pm\k\lambda
  a}=\epsilon^\mu_\lambda(\k) T^a e^{\pm ik\cdot x}$ that we have
introduced in order to perform the Fourier decomposition of classical
fields and small fluctuations. The Fourier coefficients
$h_{\pm\k\lambda a}^{(\pm)}(\p\zeta c)$ of the fluctuations
$a_{\pm\lambda a}^\mu$ can be obtained as~:
\begin{equation}
  h_{\pm\k\lambda a}^{(+)}(\p\zeta c)
  =
  -\big<{\bs a}_{-\p\zeta c}^0\big|{\bs\sigma}_2\big|{\bs a}_{\pm\k\lambda a}\big>
  \quad,\quad
  h_{\pm\k\lambda a}^{(-)}(\p\zeta c)
  =
  \big<{\bs a}_{+\p\zeta c}^0\big|{\bs\sigma}_2\big|{\bs a}_{\pm\k\lambda a}\big>
  \; .
\end{equation}
(These relations are valid only in the regions where the interactions
are switched off, i.e. when $x^0\to\pm\infty$. In the rest of the
discussion, we are only interested in these Fourier coefficients in
the limit $x^0\to +\infty$.) By multiplying eq.~(\ref{eq:projector})
by $\big<{\bs a}_{\epsilon\p\xi b}^0\big|{\bs\sigma}_2$ on the left
and by ${\bs\sigma_2}\big|{\bs a}_{\epsilon^\prime\q\zeta c}^0\big>$
on the right and using 
$\left(h_{\epsilon' \k\lambda a}^{\epsilon}(\p\zeta c)\right)^*
= h_{-\epsilon' \k\lambda a}^{-\epsilon}(\p\zeta c)$, we obtain the following relation among these Fourier
coefficients~:
\begin{eqnarray}
&&
\smash{
\sum_{\lambda,a}
\int\limits_\k
}
\Big[
h_{+\k\lambda a}^{(-\epsilon)}(\p\xi b)
h_{-\k\lambda a}^{(+\epsilon^\prime)}(\q\zeta c)
\nonumber\\
&&
\qquad\qquad\qquad
-
h_{-\k\lambda a}^{(-\epsilon)}(\p\xi b)
h_{+\k\lambda a}^{(+\epsilon^\prime)}(\q\zeta c)
\Big]
=\delta_{\epsilon\epsilon^\prime}\,\epsilon\,
(2\pi)^3\,\delta_{\xi\zeta}\delta_{bc}\, 2E_\p\delta(\p-\q)\; ,
\nonumber\\
&&
\end{eqnarray}
which is nothing but a compact way of writing the four
eqs.~(\ref{eq:unit}).

\section{Poisson distribution}
\label{app:poisson}
At first sight, eq.~(\ref{eq:Pn-final}) appears to be the average over
the distributions of sources of a Poisson distribution. This seems to
contradict a result we stressed in \cite{GelisV2}, that the
distribution of multiplicities calculated in a fixed configuration of
sources $\rho_{1,2}$ is not a Poisson distribution.  For the sake of
the discussion in this appendix, let us introduce the generating
function $F(z)$ for the multiplicity distribution in the region of
rapidity $[Y-\Delta Y/2, Y+\Delta Y/2]$. In the language of the
present paper, it is obtained by using in eq.~(\ref{eq:Fz-slice}) a
constant function $z(\p)$ whose value is equal to the number $z$.

Consider first this generating function for a given configuration
$\rho_{1,2}$ of the external color sources. In \cite{GelisV2}, $F(z)$
was parameterized as\footnote{Compared to the notations used in
  \cite{GelisV2}, we absorb the factors of $1/g^2$ into the definition
  of the numbers $b_r$.}
\begin{equation}
\ln F(z)\equiv\sum_{r=1}^\infty b_r (z^r-1)\; ,
\end{equation}
and we had obtained the formula for the probability $P_n$ of
producing $n$ particles in the portion of phase-space under
consideration to be
\begin{equation}
P_n=e^{-\sum_r b_r}\,\sum_{p=1}^n\frac{1}{p!}\sum_{r_1+\cdots+r_p=n}
b_{r_1}\cdots b_{r_p}\; .
\label{eq:Pn-generic}
\end{equation}
In Ref.~\cite{GelisV2}, we also showed that $b_r$ is the sum of all
the cut connected vacuum graphs, where exactly $r$ internal lines are
cut. Because $b_r$ is a sum of {\sl connected} graphs, it has a
perturbative expansion that starts at the order $1/g^2$,
\begin{equation}
b_r=\frac{1}{g^2}\oplus 1\oplus g^2\oplus\cdots
\end{equation}
In particular, all the $b_r$ have a priori the same order of
magnitude.  However, it is easy to see that eq.~(\ref{eq:Pn-generic})
is a Poisson distribution only in the exceptional case
where\footnote{{}From eq.~(\ref{eq:Pn-generic}) and the definition of
  the $b_r$, we have $F(z)\equiv\sum_n z^n P_n$.  Then, it is
  immediate to check that $\ln F(z)$ should be a polynomial of degree
  one in the case of a Poisson distribution.}
\begin{equation}
b_1\not=0\;,\quad b_r=0 \quad\mbox{\ for\ }r\ge 2\; .
\end{equation}
Since for a generic field theory, the $b_r$ for $r\ge 2$ have no
reason to vanish or to be smaller than $b_1$, the distribution of the
multiplicities in a fixed configuration of sources is in general not a
Poisson distribution.  Moreover, since $b_{2,3,\cdots}$ are of the same
order in $g^2$ as $b_1$, the deviations from a Poisson distribution is
an effect of order unity, not a subleading correction.

In order to make the connection with the present paper easier, it is
preferable to parameterize $F(z)$ as
\begin{equation}
\ln F(z)\equiv \sum_{k=1}^\infty c_k (z-1)^k\; .
\end{equation}
(This series starts at the index $k=1$, because $F(1)=0$.) The numbers
$c_k$ are related to the numbers $b_k$ by
\begin{equation}
b_r=\sum_{k=r}^\infty \binom{k}{r} (-1)^{k-r} c_k\; ,\quad
c_k=\sum_{r=k}^\infty \binom{r}{k} b_r\; ,
\end{equation}
where the $\binom{k}{r}$ are the binomial coefficients. The
derivatives of $\ln F(z)$ evaluated at $z=1$ are best expressed in
terms of the coefficients $c_k$ as 
\begin{equation}
\left.\frac{\partial^k \ln F(z)}{\partial z^k}\right|_{z=1}=k!c_k
\; .
\end{equation}
Let us now rephrase our results in this language.  The inclusive
$n$-particle spectrum is the $n$th derivative of $F(z)$ at $z=1$.
These derivatives read
\begin{eqnarray}
&&F^{(1)}(1)=c_1\; ,\nonumber\\
&&F^{(2)}(1)=c_1^2+2c_2\; ,\nonumber\\
&&F^{(3)}(1)=c_1^3+6 c_1 c_2+6c_3\; ,\cdots
\end{eqnarray}
All the coefficients $c_k$ are sums of connected vacuum graphs, and
therefore start at the order $1/g^2$, up to logarithms. At Leading
Order, we thus keep only
\begin{equation}
\left.F^{(n)}(1)\right|_{_{\rm LO}}
=
[c_1]_{_{\rm LO}}^n\; .
\end{equation}
At this order of truncation, one can obviously get a Poisson
distribution, since this approximation is compatible with
$c_2=c_3=\cdots=0$, i.e.  $b_2=b_3=\cdots=0$. However, the
coefficients $b_{2,3,\cdots}$ could have any value of order $g^{-2}$
without affecting our Leading Order truncation. The arbitrary choice
one is allowed to make for these subleading terms in general alters
the Poissonian nature of the distribution.

The actual paradox arises only at the Next to Leading Order. There,
one keeps the terms
\begin{eqnarray}
\left.F^{(n)}(1)\right|_{_{\rm LO+NLO}}
=
\underbrace{[c_1]_{_{\rm LO}}^n}_{g^{-2n}}
+
\underbrace{
n[c_1]_{_{\rm LO}}^{n-1}[c_1]_{_{\rm NLO}}
+
n![c_1]_{_{\rm LO}}^{n-2}[c_2]_{_{\rm LO}}
}_{g^{-2(n-1)}\times{\rm log}}\; .
\label{eq:Fn-NLO}
\end{eqnarray}
This does not correspond to a Poisson distribution anymore, since one
needs a non-zero $b_2$ in order to obtain these formulas. In fact, at
this order of truncation, one has $b_2=c_2$ while the higher $b_r$'s
are still zero. Even worse, our calculation of the second derivative
of $\ln F$ shows that $c_2$ is enhanced by a large logarithm, and is
actually of order $g^{-2}\ln(1/x_{1,2})$ rather than the naive
expectation $g^{-2}$. Therefore, not only the distribution is not
Poissonian, but the deviations from a Poisson distribution are
logarithmically large.

However, the main result of the present paper is that one can obtain
the NLO corrections to the inclusive $n$-particle spectra by the
action of a certain operator on the product of $n$ 1-particle spectra
at LO. In the present language, this reads
\begin{equation}
\left.F^{(n)}(1)\right|_{_{\rm LO+NLO}}
=
\big[
1+{\cal L}_1+{\cal L}_2
\big]\;
[c_1]_{_{\rm LO}}^n\; .
\end{equation}
Remember that so far all the discussion is for a fixed configuration
of the sources $\rho_{1,2}$. Then, by averaging over these sources and
by using the hermiticity of the operator ${\cal L}_1+{\cal L}_2$, one
can transfer the action of this operator from the quantity
$[c_1]_{_{\rm LO}}^n$ to the distribution of sources. As we have seen,
this amounts to letting the distribution of sources evolve according
to the JIMWLK equation. In other words, eq.~(\ref{eq:Fn-NLO}) deviates
strongly from a Poisson distribution, but does so in such a way that
all correlations can be interpreted as coming from correlations
among the sources that are generated by the JIMWLK evolution.

Let us end this appendix with a word of caution in the interpretation
of eq.~(\ref{eq:Fz-slice}).  Strictly speaking, our Leading Log
approximation gives us control only over the $g^{-2}\ln(1/x_{1,2})$
part of the coefficient $b_2$, but not over its $g^{-2}$ part (without
a log). The latter would only show up in a Next to Leading Log
calculation. This means that in principle one could modify the
argument of the exponential in the integrand of
eq.~(\ref{eq:Fz-slice}) by a term of second degree in $z(\p)-1$ and
with a coefficient of order $g^{-2}$, without affecting any of our
results for the inclusive gluon spectra at the order at which we
calculate them. Obviously, such a modification of the integrand in
eq.~(\ref{eq:Fz-slice}) would be a deviation from a Poisson
distribution.  Thus, the statement according to which the deviations
from Poisson come from the JIMWLK evolution of the distributions of
the sources $\rho_{1,2}$ is true only for the largest of these
deviations--i.e. those that are enhanced by large logarithms of the
momentum fractions $x_{1,2}$. Other deviations from Poisson exist,
that are not enhanced by such logarithms--these are beyond the scope
of the present calculation.

%\bibliographystyle{unsrt}
%\bibliography{biblio}

\begin{thebibliography}{10}

\bibitem{GelisLV3}
{F. Gelis, T. Lappi, R. Venugopalan}, arXiv:0804.2630 [hep-ph].

\bibitem{McLerV1}
{L.D. McLerran, R. Venugopalan}, Phys. Rev. {\bf D} {\bf 49}, 2233 (1994).

\bibitem{McLerV2}
{L.D. McLerran, R. Venugopalan}, Phys. Rev. {\bf D} {\bf 49}, 3352 (1994).

\bibitem{McLerV3}
{L.D. McLerran, R. Venugopalan}, Phys. Rev. {\bf D} {\bf 50}, 2225 (1994).

\bibitem{McLer1}
{L.D. McLerran}, Lectures given at the 40'th Schladming Winter School: Dense
  Matter, March 3-10 2001, hep-ph/0104285.

\bibitem{IancuLM3}
{E. Iancu, A. Leonidov, L.D. McLerran}, Lectures given at Cargese Summer School
  on QCD Perspectives on Hot and Dense Matter, Cargese, France, 6-18 Aug 2001,
  hep-ph/0202270.

\bibitem{IancuV1}
{E. Iancu, R. Venugopalan}, Quark Gluon Plasma 3, Eds. R.C. Hwa and X.N. Wang,
  World Scientific, hep-ph/0303204.

\bibitem{GelisLV2}
{F. Gelis, T. Lappi, R. Venugopalan}, Int. J. Mod. Phys. E {\bf 16}, 2595
  (2007).

\bibitem{JalilKMW1}
{J. Jalilian-Marian, A. Kovner, L.D. McLerran, H. Weigert}, Phys. Rev. {\bf D}
  {\bf 55}, 5414 (1997).

\bibitem{JalilKLW1}
{J. Jalilian-Marian, A. Kovner, A. Leonidov, H. Weigert}, Nucl. Phys. {\bf B}
  {\bf 504}, 415 (1997).

\bibitem{JalilKLW2}
{J. Jalilian-Marian, A. Kovner, A. Leonidov, H. Weigert}, Phys. Rev. {\bf D}
  {\bf 59}, 014014 (1999).

\bibitem{JalilKLW3}
{J. Jalilian-Marian, A. Kovner, A. Leonidov, H. Weigert}, Phys. Rev. {\bf D}
  {\bf 59}, 034007 (1999).

\bibitem{JalilKLW4}
{J. Jalilian-Marian, A. Kovner, A. Leonidov, H. Weigert}, Erratum. Phys. Rev.
  {\bf D} {\bf 59}, 099903 (1999).

\bibitem{IancuLM1}
{E. Iancu, A. Leonidov, L.D. McLerran}, Nucl. Phys. {\bf A} {\bf 692}, 583
  (2001).

\bibitem{IancuLM2}
{E. Iancu, A. Leonidov, L.D. McLerran}, Phys. Lett. {\bf B} {\bf 510}, 133
  (2001).

\bibitem{FerreILM1}
{E. Ferreiro, E. Iancu, A. Leonidov, L.D. McLerran}, Nucl. Phys. {\bf A} {\bf
  703}, 489 (2002).

\bibitem{GelisV2}
{F. Gelis, R. Venugopalan}, Nucl. Phys. {\bf A} {\bf 776}, 135 (2006).

\bibitem{GelisV3}
{F. Gelis, R. Venugopalan}, Nucl. Phys. {\bf A} {\bf 779}, 177 (2006).

\bibitem{DumitGMV1}
{A. Dumitru, F. Gelis, L. McLerran, R. Venugopalan}, arXiv:0804.3858 [hep-ph],
  to appear in Nucl. Phys. A.

\bibitem{Adamsa4}
{J. Adams, et al.}, [STAR Collaboration] Phys. Rev. Lett. {\bf 95}, 152301
  (2005).

\bibitem{Wang1}
{F. Wang}, [STAR Collaboration] talk at Quark Matter 2004, J.\ Phys.\ G {\bf
  30}, S1299 (2004).

\bibitem{Adamsa5}
{J. Adams, et al.}, [STAR Collaboration] Phys. Rev. {\bf C} {\bf 73}, 064907
  (2006).

\bibitem{Adarea1}
{A. Adare, et al.}, [PHENIX Collaboration] arXiv:0801.4545 [nucl-ex].

\bibitem{Wosie1}
{B. Wosiek}, [PHOBOS Collaboration], plenary talk at Quark Matter 2008,
  arXiv:0804.4352 [nucl-ex].

\bibitem{LappiM1}
{T. Lappi, L.D. McLerran}, Nucl. Phys. {\bf A} {\bf 772}, 200 (2006).

\bibitem{KovneMW1}
{A. Kovner, L.D. McLerran, H. Weigert}, Phys. Rev. {\bf D} {\bf 52}, 3809
  (1995).

\bibitem{KovneMW2}
{A. Kovner, L.D. McLerran, H. Weigert}, Phys. Rev. {\bf D} {\bf 52}, 6231
  (1995).

\bibitem{KovchR1}
{Yu.V. Kovchegov, D.H. Rischke}, Phys. Rev. {\bf C} {\bf 56}, 1084 (1997).

\bibitem{GyulaM1}
{M. Gyulassy, L.D. McLerran}, Phys. Rev. {\bf C} {\bf 56}, 2219 (1997).

\bibitem{KrasnV1}
{A. Krasnitz, R. Venugopalan}, Phys. Rev. Lett. {\bf 84}, 4309 (2000).

\bibitem{KrasnV2}
{A. Krasnitz, R. Venugopalan}, Phys. Rev. Lett. {\bf 86}, 1717 (2001).

\bibitem{KrasnV3}
{A. Krasnitz, R. Venugopalan}, Nucl. Phys. {\bf B} {\bf 557}, 237 (1999).

\bibitem{KrasnNV1}
{A. Krasnitz, Y. Nara, R. Venugopalan}, Nucl. Phys. {\bf A} {\bf 727}, 427
  (2003).

\bibitem{KrasnNV2}
{A. Krasnitz, Y. Nara, R. Venugopalan}, Phys. Rev. Lett. {\bf 87}, 192302
  (2001).

\bibitem{KrasnNV3}
{A. Krasnitz, Y. Nara, R. Venugopalan}, Phys. Lett. {\bf B} {\bf 554}, 21
  (2003).

\bibitem{Lappi1}
{T. Lappi}, Phys. Rev. {\bf C} {\bf 67}, 054903 (2003).

\bibitem{Lappi2}
{T. Lappi}, Phys. Rev. {\bf C} {\bf 70}, 054905 (2004).

\bibitem{Lappi3}
{T. Lappi}, Phys. Lett. {\bf B} {\bf 643}, 11 (2006).

\bibitem{KharzKV1}
{D. Kharzeev, A. Krasnitz, R. Venugopalan}, Phys. Lett. {\bf B} {\bf 545}, 298
  (2002).

\bibitem{GavinMM1}
{S. Gavin, L. McLerran, G. Moschelli}, arXiv:0806.4718 [nucl-th].

\bibitem{RomatV1}
{P. Romatschke, R. Venugopalan}, Phys. Rev. Lett. {\bf 96}, 062302 (2006).

\bibitem{RomatV2}
{P. Romatschke, R. Venugopalan}, Eur. Phys. J. {\bf A} {\bf 29}, 71 (2006).

\bibitem{Mrowc1}
{S. Mrowczynski}, hep-ph/0511052.

\bibitem{BaierKNW1}
{R. Baier, A. Kovner, M. Nardi, U.A. Wiedemann}, Phys. Rev. {\bf D} {\bf 72},
  094013 (2005).

\bibitem{JalilK2}
{J. Jalilian-Marian, Y. Kovchegov}, Phys. Rev. {\bf D} {\bf 70}, 114017 (2004),
  Erratum-ibid. {\bf D} {\bf 71}, 079901 (2005).

\bibitem{FukusH1}
{K. Fukushima, Y. Hidaka}, arXiv:0806.2143 [hep-ph].

\bibitem{FukusGM1}
{K. Fukushima, F. Gelis, L. McLerran}, Nucl. Phys. {\bf A} {\bf 786}, 107
  (2007).

\bibitem{Braun5}
{M.A. Braun}, Eur. Phys. J. {\bf C} {\bf 42}, 169 (2005).

\bibitem{Marqu1}
{C. Marquet}, Nucl. Phys. {\bf A} {\bf 796}, 41 (2007).

\bibitem{KovneL1}
{A. Kovner, M. Lublinsky}, JHEP {\bf 0611}, 083 (2006).

\bibitem{AbramGK1}
{V.A. Abramovsky, V.N. Gribov, O.V. Kancheli}, Sov. J. Nucl. Phys. {\bf 18},
  308 (1974).

\bibitem{CiafaM1}
{M. Ciafaloni, G. Marchesini}, Nucl. Phys. {\bf B} {\bf 109}, 261 (1976).

\bibitem{BarteSV1}
{J. Bartels, M. Salvadore, G.P. Vacca}, Eur. Phys. J. {\bf C} {\bf 42}, 53
  (2005).

\bibitem{BarteSV2}
{J. Bartels, M. Salvadore, G.P. Vacca}, JHEP {\bf 0806}, 032 (2008).

\end{thebibliography}

\end{document}